\documentclass[aps]{revtex4}
\usepackage{graphicx}
[1999/03/29 PSNFSS v.7.2
Times font as default roman
: S Rahtz]

\begin{document}
\renewcommand{\thefootnote}{\fnsymbol{footnote}}
\sloppy
\newcommand{\rp}{\right)}
\newcommand{\lp}{\left(}
\newcommand \be  {\begin{equation}}
\newcommand \ba {\begin{eqnarray}}
\newcommand \ee  {\end{equation}}
\newcommand \ea {\end{eqnarray}}

\title{Diffusion of epicenters of earthquake aftershock, Omori law
and generalized continuous-time random walk models}
\thispagestyle{empty}

\author{A. Helmstetter$^1$ and D. Sornette$^{2,3}$\\
$^1$ Laboratoire de G{\'e}ophysique Interne et Tectonophysique,
Observatoire de Grenoble, Universit{\'e} Joseph Fourier, BP 53X, 38041
Grenoble Cedex, France \\ e-mail: ahelmste@obs.ujf-grenoble.fr\\
$^2$ Laboratoire de Physique de la Mati\`{e}re Condens\'{e}e\\ CNRS UMR6622 and
Universit\'{e} de Nice-Sophia Antipolis\\ B.P. 71, Parc
Valrose, 06108 Nice Cedex 2, France\\
$^3$ Institute of Geophysics and
Planetary Physics and Department of Earth and Space Science\\
University of California, Los Angeles, California 90095\\
email: sornette@ess.ucla.edu}

\begin{abstract}

The epidemic-type aftershock
sequence model (ETAS) is a simple stochastic process modeling
seismicity, based on the two best-established empirical laws, the
Omori law (power law decay $\sim 1/t^{1+\theta}$ of seismicity after
an earthquake)
and Gutenberg-Richter law (power law distribution of earthquake energies).
In order to describe also the space distribution of seismicity, we use
  in addition a power law distribution $\sim 1/r^{1+\mu}$ of distances between
triggered and triggering earthquakes.
The ETAS model has been studied for the last two decades to model
real seismicity catalogs and to obtain short-term probabilistic forecasts.
Here, we present an exact mapping between the ETAS model and a class
of CTRW (continuous time random walk) models, based on the identification
of their corresponding Master equations.
This mapping allows us to use the wealth of results previously
obtained on anomalous diffusion of CTRW. After translating into the relevant
variable for the ETAS model, we provide a classification of the different
regimes of diffusion of seismic activity triggered by a mainshock.
Specifically, we derive the relation between the average distance between
aftershocks and the mainshock as a function of the time from the mainshock
and of the joint probability distribution of the times and locations of
the aftershocks.
The different regimes are fully characterized by the two exponents
$\theta$ and $\mu$.
Our predictions are checked by careful numerical simulations.
We stress the distinction between the ``bare'' Omori law describing the
seismic rate activated directly by a mainshock and the ``renormalized''
Omori law taking into account all possible cascades from mainshocks
to aftershocks
of aftershock of aftershock, and so on. In particular, we predict that
seismic diffusion or sub-diffusion occurs and should be observable only
when the observed Omori exponent is less than $1$, because this signals
the operation of the renormalization of the bare Omori law, also at the
origin of seismic diffusion in the ETAS model.
We present new predictions and insights provided by the ETAS to CTRW mapping
that suggest novel ways for studying seismic catalogs. Finally, we
discuss the present
evidence for our predicted sub-diffusion of seismicity triggered by a
main shock,
stressing the caveats and limitations of previous empirical works.

\end{abstract}

\maketitle

\pagenumbering{arabic}

\section{Introduction}

The spatio-temporal complexity of earthquakes is often invoked as
an illustration of the phenomenon of critical self-organization with
scale-invariant properties
\cite{Geilo,Rundleklein,Main,Sorreview,Turcottereview}.
This concept points to the importance of developing a system approach in which
large scale properties can emerge from the repeating interactions occurring
at smaller scales. Such ideas are implemented
in models proposing links between the physics of earthquakes
and concepts of statistical physics, such as
critical points, self-organized criticality, spinodal decomposition,
critical depinning, etc., in order to explain the most
solidly established facts in
the phenomenology of earthquakes, of which we cite the three most important.
\begin{itemize}
\item LAW 1: The Gutenberg-Richer law \cite{GR} states that the cumulative
distribution of earthquake magnitudes $m$
sampled over broad regions and large time intervals is proportional
to $10^{-b m}$, with a $b$-value $b \approx 1$. Translating into energies $E$
with the correspondence
$m=(2/3) \log_{10} E + $ constant leads to a
power law $\sim 1/E^{B}$ with $B \approx 2/3$.

\item LAW 2: Omori law for aftershocks \cite{Omori} states that the
rate of earthquakes
triggered by a mainshock decays with time according to an inverse power
$1/t^{p}$ of time with an exponent $p \approx 1$.
\item LAW 3: The earthquakes are clustered in space along hierarchical fault
structures \cite{Ouillonetal} and their spatial distribution over
long times can be approximately
described by a fractal dimension close to $2.2$ (in three dimensions)
\cite{KK80}.
\end{itemize}
There are many other empirical ``laws'' but these three characterize the
very fundamentals of seismicity in size, time and space.

We should immediately point out that these three laws come with
significant caveats.
\begin{enumerate}
\item There have been on-going controversies on the universality
of the exponent $B$ or $b$-value of the Gutenberg-Richter law \cite{Pissor,kaganuniv}.

\item The exponent $p$ of Omori law exhibits a large variability from one
aftershock sequence to another aftershock sequence and
is found typically in the range from $0.3$ to $2$.
We note however that not all these values, especially the extreme ones,
automatically reflect a bona-fide power law decay and one should exert
caution in attributing too much confidence to them.

\item The view that geological faults and earthquake hypocenters are fractal
objects is now recognized to be a naive description of a much more
complex reality in which a hierarchy of scales occur with possibly
different organizations at different scales \cite{Ouillonetal}.
\end{enumerate}

In addition, a major difficulty for making progress in modeling and
predicting earthquakes
is that these three and other laws may be ``explained'' by a large variety of
models, with many different mechanisms. For instance,
with respect to the first two laws, we observe the
following.
\begin{itemize}
\item There are many mechanisms that create a power
law distribution of earthquake sizes (see for instance the list of
mechanisms described in chapter 14 of Ref.~\cite{Sorbook}).
\item Omori law
is essentially a slowly decaying ``propagator'' describing
a long-time memory of past events impacting on the
future seismic activity. Such slow power law time decay of the
Omori propagator
may result from several and not necessarily exclusive mechanisms
(see \cite{Harrisreview}
and references therein): pore-pressure changes due to pore-fluid flows coupled
with stress variations, slow redistribution of stress by aseismic creep,
rate-and-state dependent friction within faults, coupling between
the viscoelastic lower crust and the brittle upper crust, stress-assisted
micro-crack corrosion \cite{Yamakno,Leesor}, slow tectonic driving of
a hierarchical geometry
with avalanche relaxation dynamics \cite{Huangetal}, etc.
\end{itemize}

The zeroth order description of earthquakes is to consider a single isolated
homogeneous fault on which earthquakes are recurrent to accommodate
the long-term slow tectonic loading. But faults are not isolated and
the most conspicuous observation is that earthquakes interact and
influence each other on complex fault structures. Understanding these
interactions is essential
for understanding earthquakes and fault self-organization. However, the
full impact of interactions between earthquakes is still far
from being well understood. The simplest and clearest observation of
earthquake interaction is provided by aftershocks whose phenomenology
is captured by Omori law (LAW 2). Indeed,
aftershocks are the most obvious and striking signature of the clustering
of the seismicity in time and space, and are observed after all
large  shallow earthquakes.
Most aftershocks are triggered a few hours or days after the mainshock.
However, due to the very slow power law decay of the rate of aftershock,
known as the Omori law \cite{Omori}, aftershocks can be triggered
up to a hundred years after the mainshock \cite{Utsu}.
Aftershocks often occur near the rupture zone of the mainshock
with a variety of focal mechanisms suggesting
that they are actually on separate structures 
\cite{BathRichter,Beroza}. They are also sometimes
triggered at very large
distances from the mainshock \cite{Hill,Steeples,KJ1,Meltzner1,Dreger}.
As an example, Hill et al. \cite{Hill} observed aftershocks
of the Landers earthquake as far as 1250 km  from the epicenter.
Similarly to the temporal distribution of aftershocks, a power-law distribution
seems to describe well the distribution of distances between pairs
of events \cite{KJ1} .
Since a power-law decays slowly, it describes a slow decay of the
probability of
observing aftershocks  at large distances to the mainshock.

Thus, Omori law can be considered as the simplest and best established
description of earthquake interactions of a certain kind. The question we
investigate is whether it can be used fruitfully to explain a larger
variety of earthquake interactions beyond the class of observations that
were used to establish it. In a series of papers \cite{SS,HS1,SH2},
we find that
Omori law for aftershocks plus the constrain that aftershocks
are distributed according to the Gutenberg-Richter power law for
earthquake size distribution {\it independently} of the magnitude of their
progenitor is enough to derive many of the other empirical ``laws,'' as well
the variability of the $p$-exponent.
Here, we test the potential of this approach to account for
and to quantify observations on aftershock diffusion.

Aftershock diffusion refers to the phenomenon of
expansion or migration of aftershock zone
with time
\cite{Mogi,Imoto,Chatelain,Tajima1,Tajima2,Wesson,Ouchi,Noir,Jacques}.
Immediately after the
mainshock occurrence, most aftershocks are located close to the rupture
plane of the mainshock, then aftershocks seem to migrate away from
the mainshock,
at velocity ranging from 1 km/hour to 1 km/year \cite{Jacques,Rydelek}.
Note that this expansion is not universally observed,
but is more important in some areas than in others \cite{Tajima1}.

The diffusion of aftershocks is usually interpreted as a diffusion of
the stress
induced by the mainshock, either by a viscous relaxation process
\cite{Rydelek},
or due to fluid transfer in the crust \cite{Nur,Hudnut,Noir}.
Another interpretation of the expansion of aftershocks is given by
Dieterich \cite{Diete},
who reproduces the Omori law decay of aftershocks and the expansion of the
    aftershock zone with time, using a rate and state friction law and
assuming that the
    rate of aftershocks is proportional to the stress rate.
In his model, the expansion of aftershock zone arises from the
non-uniform stress
induced by the mainshock.
Another alternative explanation is
that the diffusion of aftershocks is mainly due to the occurrence
    of large aftershocks, and to the localization of secondary
aftershock close to
the largest aftershocks, as observed by Ouchi \cite{Ouchi}.
The apparent diffusion of the seismicity may thus result from a
cascade process;
the mainshock triggers aftershocks that in turn trigger their own
aftershocks, and thus lead to an expansion of the aftershock zone.

In the present paper, we investigate the epidemic time aftershock
sequence model (ETAS),
and show that the cascade of secondary aftershocks can indeed explain
the reported diffusion of aftershocks.
The ETAS model was introduced by Kagan and Knopoff \cite{KK} 
(in a slightly different form than used here) and Ogata \cite{Ogata1}
to describe the temporal and spatial clustering of seismicity.
This model provides a tool for understanding the
clustering of the seismic activity, without arbitrary distinction between
aftershocks, foreshocks and mainshocks.
In this model, all earthquakes are assumed to be simultaneously
mainshocks, aftershocks and possibly foreshocks. Each earthquake generates
aftershocks that decay with time according to Omori law, which
will in turn generate their own aftershocks. The seismicity rate at
any given time and location is given by
the superposition of aftershock sequences of all events impacting that region
at that time according to space-time ``propagators.''
The additional ingredient in the version of the ETAS model that we
study is that the number of aftershocks per earthquake increases exponentially
$\propto 10^{\alpha m}$ with the magnitude $m$ of
the mainshock (i.e., as a power law $\propto E^{2 \alpha /3}$
of the energy released by the mainshock), in agreement with the observations
\cite{YaShi,Drakatos}.
Since the energy of an earthquake is a power
law of its rupture length, this law expresses
the very reasonable idea that the number of events related to a given
earthquake is proportional to a power of its volume of influence.
The value of the exponent $\alpha$ controls the nature of
the seismic activity, that is, the relative role of small compared
to large earthquakes. Few studies have measured $\alpha$
in seismicity data \cite{YaShi,Guo2,H02}.
This parameter $\alpha$ is often found close to $b$ \cite{YaShi}
 or fixed arbitrarily equal to $b$ \cite{KK,Felzer}.
In the case where $\alpha$ is close to the Gutenberg-Richter $b$-value,
this law also reproduces
\cite{Felzer} the self-similar empirical Bath' s law \cite{Bath},
which states that the
average difference $m_M-m_A$ in size between a mainshock and its largest aftershock is
$1.2$ magnitude units, regardless of the mainshock magnitude:
$m_A = m_M - 1.2$.
If $\alpha < b$, small earthquakes, taken together, trigger more
aftershocks than larger
earthquakes. In contrast, large earthquakes dominate earthquake
triggering if $\alpha \geq b$.
This case $\alpha \geq b$ has been studied analytically in the
framework of the ETAS model by Ref. \cite{SH2}
and has been shown to eventually lead to a finite time singularity of
the seismicity rate. This explosive regime cannot however describe
a stationary seismic activity. 

A natural way to tame this singular behavior is to introduce an
upper cut-off for the magnitude distribution at large magnitudes,
mirroring the cut-off $m_0$ used for the low-magnitude range. The
physical argument for introducing this cut-off is based on the finiteness 
of the maximum earthquake that the earth is capable of carrying.
The specific way of introducing such a cut-off (abrupt or smooth
with a transition to a power law with larger exponent or to an exponential taper) is not
very important qualitatively because all these laws will regularize the singular
behavior and make the average branching ratio finite. Such 
regularization with a maximum upper magnitude then allows $\alpha \geq b$.
The special case $\alpha = b$ required for Bath's law to hold exactly
can not therefore be excluded.

However, based on a recent re-analysis of seismic catalogs using the powerful
collapse technique, one of us \cite{H02} has presented strong evidence
that $\alpha$ is strictly smaller than $b$.
In this paper, we will therefore consider only the case $\alpha<b$
and take $\alpha=0.5$ specifically in our numerical simulations.
In this regime $\alpha<b$, Bath' s law cannot be reproduced because
the average difference in size between a mainshock and its largest aftershock
increases with the mainshock magnitude. For $\alpha < b$, it is 
easy to show that Bath's law
is replaced by $m_A = (\alpha/b) m_M -$ constant, where $m_M$ and 
$m_A$ are the magnitudes of the mainshock and of the largest aftershock.
Tests of this prediction will be reported in a future publication but 
we expect that distinguishing this
modified Bath's law from Bath's law will be a difficult task due to the
limited range of the studied magnitudes as well as the dependence 
of the distribution of $m_M-m_A$ on the magnitude thresholds chosen for the mainshocks and
for the aftershocks \cite{consolebath}.

We assume that the distribution of all earthquakes follow the
Gutenberg-Richter distribution and take this
distribution of aftershock sizes to be independent of the magnitude
of the mainshock. Therefore,
an earthquake can trigger a larger earthquake, albeit with
a small probability. This model can thus describe a priori
both aftershock and foreshock sequences.
The ETAS model has been calibrated to real seismicity
catalogs to retrieve its parameters
\cite{Ogata1,Ogata2,Ogata3,Ogata5,Ogata6,Kagan,Guo2,Felzer}
and to give short-term probabilistic forecast of
seismic activity by extrapolating past seismicity into the future
via the use of its space-time propagator \cite{KK,KJ2,Console}.

The ETAS model is a branching model which exhibits different
regimes \cite{HS1} depending upon the value of
the branching ratio $n$, defined by the
average number of primary aftershocks per earthquake.
The critical case $n=1$ corresponds to exactly one primary aftershock
per earthquake, when averaging over all mainshock magnitudes larger
than a threshold $m_0$. Let us stress that $n$ is an average quantity
which does not reflect adequately the large variability of the number of
aftershocks per main shock, as a function of its magnitude. Indeed,
the number of aftershocks per mainshock
increases exponentially fast as a function of the mainshock magnitude,
so that large mainshocks
will have significantly more than $n$ aftershocks. For $\alpha =0.5$,
 a magnitude $7$-earthquake gives typically $10$
times more direct aftershocks than a magnitude $5$, and $100$ times more
direct aftershocks than a magnitude $3$-earthquake. The increase
in triggered seismic activity with the magnitude of the mainshock
is obviously stronger for a larger value of $\alpha$. Note that these
numbers refer to aftershocks of the first generation; the total number
of triggered events is larger by the factor $1/(1-n) \sim 10$ 
(for $n \approx 0.9$ which is typical),
due to the cascades of secondary aftershocks.
Notwithstanding this large variability, the average
number $n$ of primary aftershocks per earthquake controls the global regime.
For $n$ exactly equal to $1$,
seismicity is at the border between death and growth.
In the sub-critical regime $n<1$, since each earthquake triggers on
average less
that one aftershock, starting from a large event, the seismicity will decrease
with time and finally die out.
The super-critical  $n>1$ corresponds to more that one primary aftershock per
earthquake on average. Starting from a large earthquake, after a transient
regime, the average seismicity will finally increase exponentially 
with time \cite{HS1},
but there is still a finite probability for aftershock sequences to die out.

The numerical simulations reported below have been performed with $\alpha=0.5$.
It is probable that a good fit to seismic
data is obtained by using a value of $\alpha \approx 0.8$ larger that the value $0.5$,
as reviewed and documented recently by one of us \cite{H02}.
We have checked that results similar to those presented below hold true qualitatively for larger 
values $0.5< \alpha <1$. Such larger values of $\alpha$ lead however to stronger fluctuations
which are more difficult to handle numerically because the variance of
the number $\rho(m)$ of direct triggered aftershocks defined below in 
(\ref{formrho}) becomes undefined for $\alpha > 0.5$. 
A full understanding of this regime requires a special treatment
that will be reported elsewhere.

Sornette and Sornette \cite{SS} studied analytically a particular case of this
model, without magnitude and spatial dependence, and they considered only the
subcritical regime $n<1$. Starting with one event at time $t=0$ and
considering that each earthquake generates an aftershock sequence
with a ``local''   Omori exponent $p=1+\theta$, where $\theta>0$,
they studied  the decay law of the ``global'' aftershock sequence,
composed of all secondary  aftershock sequences, i.e., by taking into
account that the primary aftershocks can create secondary aftershocks
which themselves may trigger tertiary aftershocks and so on.
They found that the global aftershock rate decays according to an Omori law
with an exponent $p=1-\theta <1$,
up to a characteristic time \cite{SS,HS1}
\be
t^* = c ~\left({n~\Gamma(1-\theta) \over |1-n|}\right)^{1/\theta}~,
\label{tstar}
\ee
and then recovers the local Omori exponent
$p=1+\theta$ for time larger than  $t^*$.
Helmstetter and Sornette \cite{HS1} extended their analysis to the general ETAS
model with magnitude dependence, and considered both the sub- and the
super-critical regime, but still restricted the analysis to the temporal
distribution of the seismicity, without spatial dependence.
 In the sub-critical regime, they recovered
the crossover found by Sornette and Sornette \cite{SS}. 
In addition, Helmstetter and Sornette \cite{HS1} give the explicit mathematical
formula for the gradual transition between the Omori law with exponent 
$p=1-\theta$ for $t \ll t^*$ to the Omori law with exponent
$p=1+\theta$ for $t \gg t^*$. This smooth transition can be observed
in figure \ref{n} on the line calculated for $t^*=10^9$ days 
with $n<1$. $t^*$ can thus be viewed as the time where the 
apparent exponent $p$ of the Omori law is
approximately in between the two asymptotic values $1-\theta$
and $1+\theta$. A more rigorous
mathematical definition \cite{HS1} is that $t^*$ is the
characteristic time scale such that $\beta t^*$ is the dimensionless
variable of the Laplace transform (with variable $\beta$) of the seismicity rate. 

In the super-critical
regime, Helmstetter and Sornette \cite{HS1}
found a novel transition between a power-law decay with exponent
$p=1-\theta$ at early times, similar to the sub-critical regime, to
an exponential increase of the seismicity at large times.
The regime where $\alpha > b$ or equivalently $2 \alpha/3 > B$ has been found
to lead to a new kind of critical stochastic finite-time-singularity
\cite{SH2}, relying on the interplay
between long-memory and extreme fluctuations. Recall that
the number of aftershocks per earthquake increases as a power law 
$\propto E^{2 \alpha/3}$ of the energy released by the mainshock
whereas the number of earthquakes of energy $E$ decreases as
the Gutenberg-Richter law $\propto 1/E^{1+B}$.
Intuitively, when $2 \alpha /3 > B$,
the increase in the rate of creation of aftershocks with the mainshock
energy more than compensate the decrease of the probability to get
a large mainshock when the mainshock energy increases.
This theory based solely on the ETAS model has been found to account for the
main observations (power law acceleration and discrete scale invariant
 structure) of critical rupture of heterogeneous materials, of the largest
sequence of starquakes ever attributed to a neutron star as well as of some
earthquake sequences \cite{SH2}.

In the sequel, we extend the analytical study of the temporal ETAS
model \cite{SS,HS1,SH2}
to the spatio-temporal domain. To model the spatial distribution of
aftershocks, we assume that the distance between a mainshock and each
 of its direct aftershock is drawn from a given distribution,
 independently of the magnitude of the mainshock and of
the delay between the mainshock and its aftershocks. For illustration but
without loss of generality for the mapping to the
continuous time random walks model (CTRW) discussed later, we
shall take a power law distribution of distances between earthquakes.
We take the simplest and most parsimonious hypothesis that
space, time and magnitude are decoupled in the earthquake propagator.
Our first result is to establish a correspondence between
the ETAS model and the CTRW, first introduced
by Montroll and Weiss \cite{Montroll1} and used to model many
physical processes.
We then build on this analogy to derive the joint
probability distribution of the times and locations of aftershocks.
We show analytically that, for sufficiently short times $t<t^*$,
the average distance between a mainshock and its aftershock increases
subdiffusively
as $R \sim t^H$, where the exponent $H$ depends on the local Omori
exponent $1+\theta$ and
on the distribution of the distances between an earthquake and its aftershocks.
We also demonstrate that the local Omori law is not universal, but varies as
a function of the distance from the mainshock. Due to the diffusion
of aftershocks with time, the decay of aftershock is faster close
to the mainshock than at large distances. These non-trivial space-time
couplings occur notwithstanding the decoupling
between space, time and magnitude in the ``bare'' propagator, and is due to
the existence of cascades of aftershocks.

A recent work of Krishnamurthy et al. \cite{Tanguy} substantiates the general
modeling strategy used here of representing the space-time dynamics of earthquakes
by an effective stochastic process (the ETAS model) entirely defined
by two exponents
(corresponding to our $\mu$ and $H(\theta, \mu)$ defined below), where $\mu$ is
the exponent of
the power law distribution of jumps between successive active sites and $H$
is the (sub-)diffusion exponent.
Indeed, Krishnamurthy et al. \cite{Tanguy} show
that the Bak and Sneppen model and the Sneppen model of extremal dynamics
(corresponding to a certain class of self-organized critical behavior
\cite{Sorbook})
can be completely characterized by a suitable stochastic process called
``Linear fractional stable motion.'' Beyond recovering the scaling exponents
of this model, the stochastic process strategy predicts the conditional
probabilities of successive activations at different sites and thus offers
novel insights. We note that this approach with the
Linear fractional stable motion is extremely close in spirit as well as
in form to our approach mapping the ETAS model to the CTRW model.
The ETAS model can thus be taken to represent an effective stochastic process
of the complex self-organization of seismicity.

\section{The ETAS model}

\subsection{Definitions and specific parameterization of the ETAS model}

We assume that a given event (the ``mother'') of magnitude
$m_i$ occurring at time $t_i$ and position $\vec r_i$
gives birth to other events (``daughters'') of any possible
magnitude chosen with some independent Gutenberg-Richter
distribution at a later time
between $t$ and $t+dt$ and at point $\vec r \pm \vec dr$ to within
$d\vec r $ at the rate
\be
\phi_{m_i}(t-t_i, \vec r-\vec r_i) = \rho(m_i)~\Psi(t-t_i)~\Phi(\vec
r-\vec r_i)~.
\label{first}
\ee
We will refer to $\phi_{m_i}(t-t_i, \vec r-\vec r_i)$ both as
the seismic rate induced by a single mother or as the ``bare propagator''.
It is the product of three independent contributions:
\begin{enumerate}
\item $\rho(m_i)$ gives the number of daughters born from a mother with
magnitude $m_i$. This term will in general be chosen to account for the
   fact that large earthquakes have many more triggered events than small 
earthquakes. 
Specifically, we take
\be
\rho(m_i) = K ~10^{\alpha (m_i-m_0)}~,  \label{formrho}
\ee
which, as we said earlier, is justified by the power law dependence of the
volume of stress perturbation as a function of the earthquake size. 
$\alpha$ quantifies how fast the
average number of daughters per mother increases with the magnitude of the mother.

\item $\Psi(t-t_i)$ is a normalized waiting time distribution giving the rate
of daughters born at time $t-t_i$ after the mother. The normalization condition
reads $\int_0^{+\infty} dt ~\Psi(t) =1$. $\Psi(t-t_i) dt$  can thus
be interpreted as the probability for a daughter to be born between $t$ and
 $t+dt$ from the mother that was born at time $t_i$. $\Psi(t-t_i)$ embodies
  Omori law: it is the ``bare'' or ``direct'' Omori law
\be
\Psi(t) = {\theta~c^{\theta}  \over (t+c)^{1+\theta}}~H(t)~,
\label{psidef}
\ee
where $\theta>0$ and $H(t)$ is the Heaviside function.

\item $\Phi(\vec r-\vec r_i)$ is a normalized spatial ``jump'' distribution from
the mother to each of her daughter, quantifying the probability for a daughter
to be triggered at a distance $|\vec r-\vec r_i|$ from the mother.
Specifically, we take
\be
\Phi(\vec r) = {\mu \over d ({|\vec r| \over d}+1)^{1+\mu}}~,
\label{phidef}
\ee
which has the form of an (isotropic) elastic Green function
dependence describing the stress transfer in an elastic upper crust.
The exponent $\mu$ is left adjustable to account for heterogeneity
and the possible complex modes of stress transfers.
The normalization condition reads $\int d\vec r ~\Phi(\vec r) =1$ where
the integral is carried out over the whole space.
\end{enumerate}

The physical justification for this decoupled model (\ref{first}) in which
$\phi_{m_i}(t-t_i, \vec r-\vec r_i)$ is the product of three independent
distributions is that elastic waves propagate at kilometers per second
and thus almost instantaneously reset the stress field after a large
main shock.
In other words, there is a well-defined separation of time scales
between the time
of propagation of seismic waves (seconds to minutes)
which control the convergence to a new mechanical
equilibrium after the main shock and the time scales involved in
aftershock sequences
(hours, days, months to many years).
The spatial dependence in (\ref{first}) reflects the stress redistribution.
This new stress field then relaxes slowly and more or less
independently from point to
point leading to the local Omori law $\Psi(t-t_i)$. Notwithstanding this
argument, the decoupling in (\ref{first})
between the local responses in magnitudes, space and time
is mostly performed because of its simplicity. It constitutes
an approximation that should be checked and relaxed in future
studies.

We assume a distribution $P(m)$ of earthquake
sizes expressed in magnitudes $m$ which follows the Gutenberg-Richter
distribution
\be
P(m) = b~ \ln(10) ~ 10^{-b (m-m_0)}~, \label{gojfwo}
\ee
with a $b$-value usually close to $1$.
   $m_0$ is a lower bound magnitude below which no daughter is triggered.

\subsection{The branching ratio $n$}

A key parameter of the ETAS is the average number $n$ of
daughter-earthquakes
created per mother-event, summed over all possible magnitudes. As we shall see,
it is also natural to call it the ``branching ratio''.
To see this,
consider the integral of the seismic rate $\phi_{m_i}(t-t_i, \vec r-\vec r_i)$
induced by one earthquake
over all times after $t_i$, over all spatial positions and over all
magnitudes $m_i \geq m_0$,
which must give by definition the average number $n$ of direct
(or primary) daughter-earthquakes
created per mother-event independently of its magnitude.
For $\alpha<b$ and using (\ref{first}), (\ref{formrho}) and (\ref{gojfwo}),
it is exactly given by
\be
n \equiv \int d\vec r ~ \int_{t_i}^{+\infty} dt ~\int_{m_0}^{+\infty}
dm_i~P(m_i)~
\phi_{m_i}(t-t_i, \vec r-\vec r_i) =
\int_{m_0}^{+\infty} dm_i~P(m_i)~ \rho(m_i)={K~b\over b-\alpha} ~,
\label{second}
\ee
since the two integrals over time and space contribute
each a factor $1$ by the normalization of $\Psi$ and $\Phi$.
This result (\ref{second}) is identical to that found
in absence of spatial dependence of
$\phi_{m_i}(t-t_i)$ with respect to $\vec r-\vec r_i$ due to the
factorization of the rate $\rho$,
time $\Psi$ and space $\Phi$ dependences \cite{HS1}.
The branching ratio has also been evaluated in the  case
where the magnitude distribution follows a gamma distribution \cite{Kagan}.

We stress again that $n$ is an {\it average} quantity which does not reflect
the large fluctuations in the number of aftershocks from events to events.
Indeed, large events with magnitudes $M$ produce
in general many more aftershocks than small events with magnitude $m<M$, simply
because $\rho(M) \gg  \rho(m)$ if $M>m$ (see the exponential
dependence (\ref{formrho}) of $\rho(m)$ on the magnitude $m$).

\subsection{Numerical simulation of the spatial ETAS model}
\label{num}
The ETAS model has been simulated numerically using the algorithm
described in Refs.~\cite{Ogata4,Ogata5}.
Starting with a large event of magnitude $M$ at time $t=0$, events are
then simulated sequentially. At any given time $t$, we calculate the
conditional seismic rate $\lambda(t)$ defined by
\be
\lambda(t)=\sum_{t_i\leq t} K~10^{\alpha(m_i-m_0)}
 {\theta c^\theta \over (t-t_i+c)^{1+\theta}}
\label{mgmkr}
\ee
where $K = n  (b-\alpha)/b$, and $t_i$ and $m_i$ are the
times and magnitudes of all preceding events that occurred at time $t_i\leq t$.
Note that we use the bare propagator because the sum in (\ref{mgmkr}) is
performed exhaustively on the complete catalog of past events.
The time of the following event is then determined according to the
non-stationary Poisson process of conditional intensity $\lambda(t)$,
and its magnitude is chosen according to the Gutenberg-Richter
distribution with parameter $b$.
To determine the position in space of this new event, we first choose
its mother randomly among all preceding events with a probability proportional to
their rate of aftershocks $\phi_{m_i}(t-t_i)$ evaluated at the time
of the new event. Once the mother has been chosen, we generate the
distance $r$ between the new earthquake and its mother according to
the power-law distribution $\Phi(\vec r)$ given by (\ref{phidef}).
The location of the new event is determined by assuming an isotropic
distribution of aftershocks.
By this rule, it is clear that new events tend to be close in general
to the last large earthquakes, leading to space clustering.

Note that this two-steps procedure is equivalent to
but more convenient for a numerical implementation than the one-step method,
consisting of calculating at each point on a fine
space-covering grid the seismic rate, equal to the sum over all
preceding mothers
weighted by the bare space $\Phi(\vec r)$ and time $\Psi(t)$ propagators
given by (\ref{phidef}) and (\ref{psidef}); after normalizing, these rates then
provide to each grid point a probability for the event to occur on that point.
The equivalence between our two-step procedure and the direct calculation
of the seismic rates is based on the law of conditional probabilities:
probability of next event ($A$) = probability of next event conditioned on its mother
(event $B$) $\times$ probability of choosing the mother, i.e., $P(A,B) = P(A | B)
\times P(B)$.

\begin{figure}
\includegraphics[width=16cm]{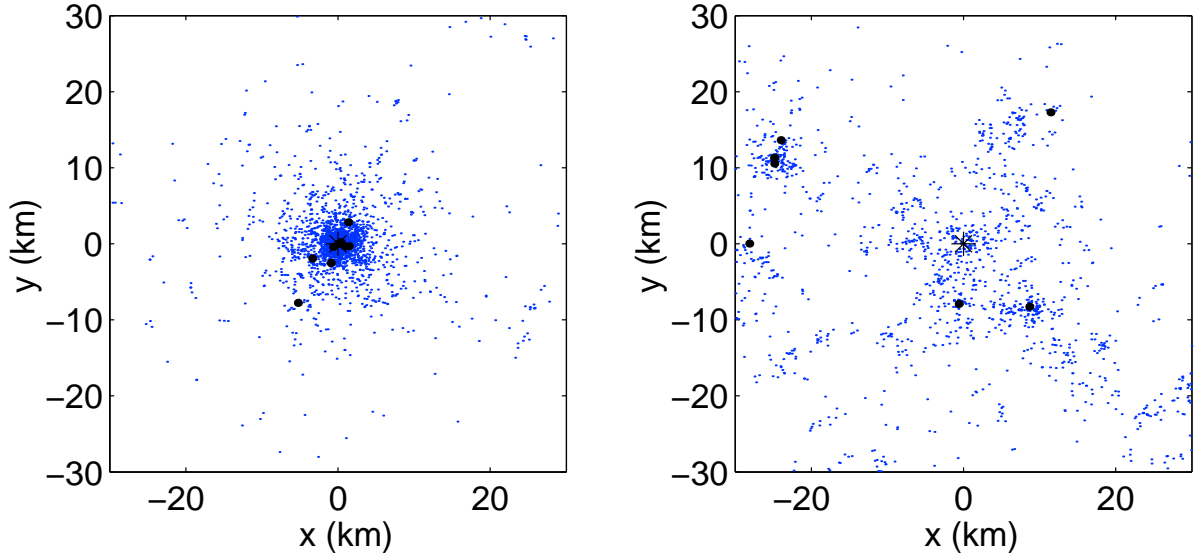}   
\caption{\label{map} Maps of seismicity generated by the ETAS model
with parameters
$b=1$, $\theta=0.2$, $\mu=1$, $d=1$ km, $\alpha=0.5$, $c=0.001$ day and a
branching ratio $n=1$.
The mainshock occurs at the origin of space with magnitude $M=7$. The minimum
    magnitude is fixed at $m_0=0$.
The distances between mainshock and aftershocks follow a power-law
with parameter
$\mu=1$ and the local (or bare) Omori law is $\propto 1/t^{1+\theta}$.
  According to the theory developed
in the text, the average distance between the first
mainshock and the aftershocks is thus expected to grow as $R \sim t^{H}$
with $H=0.2$ (equation (\ref{mgjgkr})).
The two plots are for different time periods of the same numerical
simulation, such that the same number of earthquakes $N=3000$ is obtained
for each graph: (a) time between $0$ and $0.3$ days;
(b) time between $30$ and $70$ yrs.
Real aftershock sequences are indeed observed to last decades 
up to a century. Large black dots indicate large
aftershocks around which other secondary aftershocks cluster.
The mainshock is shown by a black star. At early times, aftershocks
are localized close to the mainshock, and then diffuse and cluster around the
largest aftershocks.
}
\end{figure}

Figure \ref{map} shows the result of a numerical simulation of the ETAS model
which exhibits a diffusion of the seismic activity.
We simulate a sequence of aftershocks and secondary aftershocks starting
from a mainshock of magnitude $M=7$, with the following parameters:
$\theta=0.2$, $b=1$, $\alpha=0.5$, $n=1$ and $\mu=1$.
At early times, aftershocks are localized close to the mainshock,
and then diffuse and cluster close to the largest aftershocks. This
(sub-)diffusion is extremely slow, as we shall quantify in the sequel.
Our purpose is to provide a theory for this process based on the ETAS
model. This theory will be tested by numerical simulations.

\begin{figure}
\includegraphics[width=16cm]{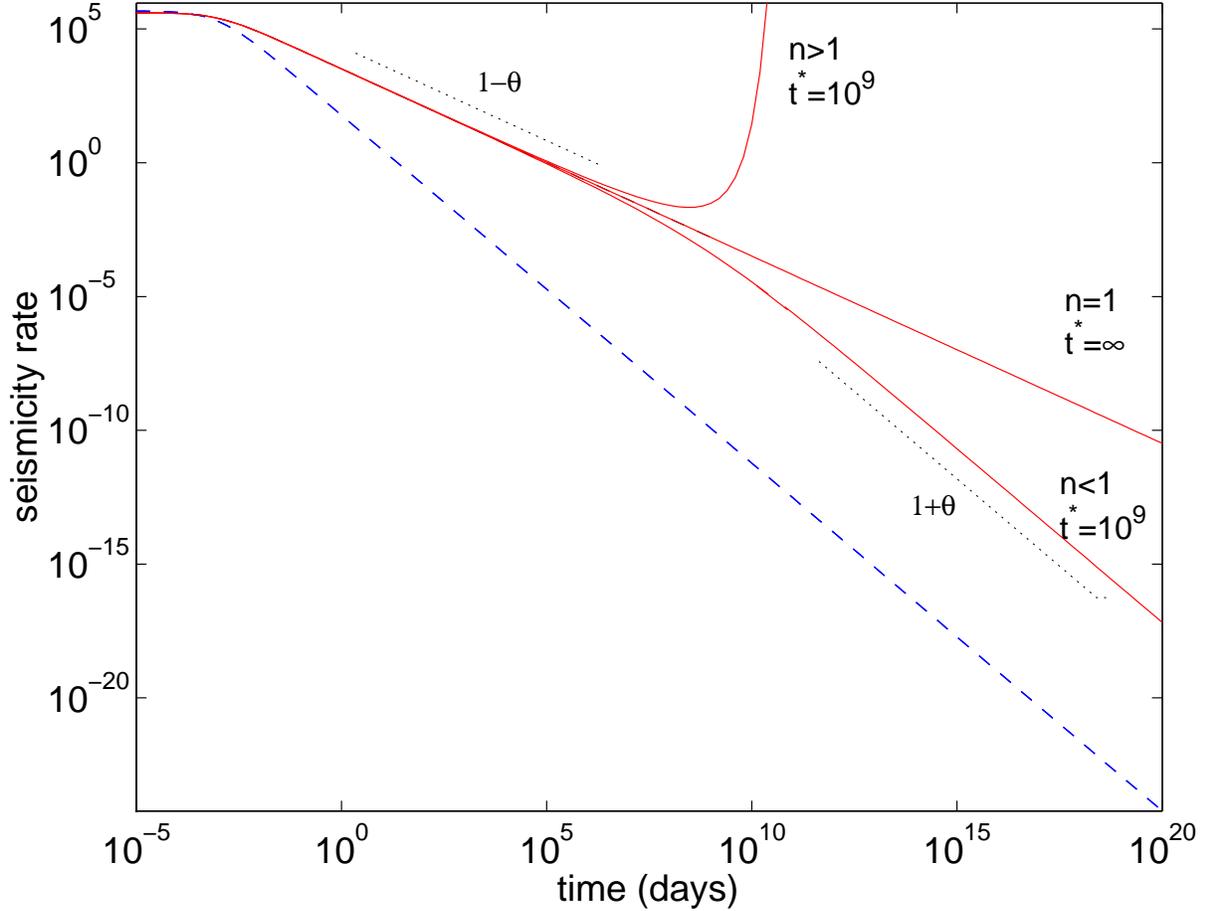}
\caption{\label{n} Seismicity rate $N(t)$ for the temporal ETAS model
calculated  for $\theta=0.3$ and $c=0.001$ day. The local law $\phi(t) \propto
1/t^{1+\theta}$, which gives the
   probability distribution of times between an event and its
(first-generation) aftershocks is
    shown as a dashed line.
   The global law $N(t)$, which includes all secondary and successive
aftershocks generated
    by all the aftershocks of the first event, is shown as a solid line for the
     three regimes, $n<1$, $n=1$ and $n>1$.
     In the critical regime $n=1$, the seismicity rate follows a
renormalized or dressed
Omori law $\propto 1/t^p$ for $t>c$ with an
exponent $p=1-\theta$, smaller than the exponent of the local law $1+\theta$.
     In the sub-critical regime ($n<1$), there
     is a crossover from an Omori law $1/t^{1-\theta}$ for $t<t^*$
     to $1/t^{1+\theta}$ for $t>t^*$. In the super critical regime
($n>1$), there
     is a crossover from an Omori law $1/t^{1-\theta}$ for $t<t^*$ to an
      exponential increase $N(t) \sim \exp(t/t^*)$ for $t>t^*$. We have chosen
      on purpose values of $n=0.9997 <1$ and $n=1.0003 > 1$ very close to $1$
      such that the crossover time $t^* =10^9$ days given by (\ref{tstar}) is
      very large. In real data, such large $t^*$ would be undistinguishable from
      an infinite value corresponding to the critical regime $n=1$.
This representation
      is chosen for pedagogical purpose to make clear the different
regimes occurring
      at times smaller and larger than $t^*$. In reality, we can expect $n$ to
      be significantly smaller or larger than $1$, such that $t^*$ becomes
      maybe of the order of months, years to decades and the observed
Omori law
      will thus lie in the cross-over regime, given an apparent Omori exponent
      anywhere from $1-\theta$ to $1+\theta$.
      }
    \end{figure}

The different regimes are illustrated by Figure \ref{n} which shows
the seismicity rate $N(t)$ for the temporal ETAS model studied by 
\cite{SS,HS1} obtained by summing the seismic activity over all space
for the 3 cases $n<1$ (sub-critical), $n=1$ (critical) and $n>1$ 
(super-critical). 
The sub-critical regime is characterized by the existence of
the time scale $t^*$ given by (\ref{tstar}).
There is no difference between the critical case $n=1$ and the
sub-critical case for $t<t^*$ (see figure \ref{n}). Indeed, the difference
between the sub-critical regime and the critical regime can be observed
only for $t>t^*$. A simple way to see this is to realize that the critical
regime $n=1$ gives $t^* = +\infty$, meaning that, in the critical regime,
one is always in the situation $t<t^*$.

\begin{figure}
\includegraphics[width=16cm]{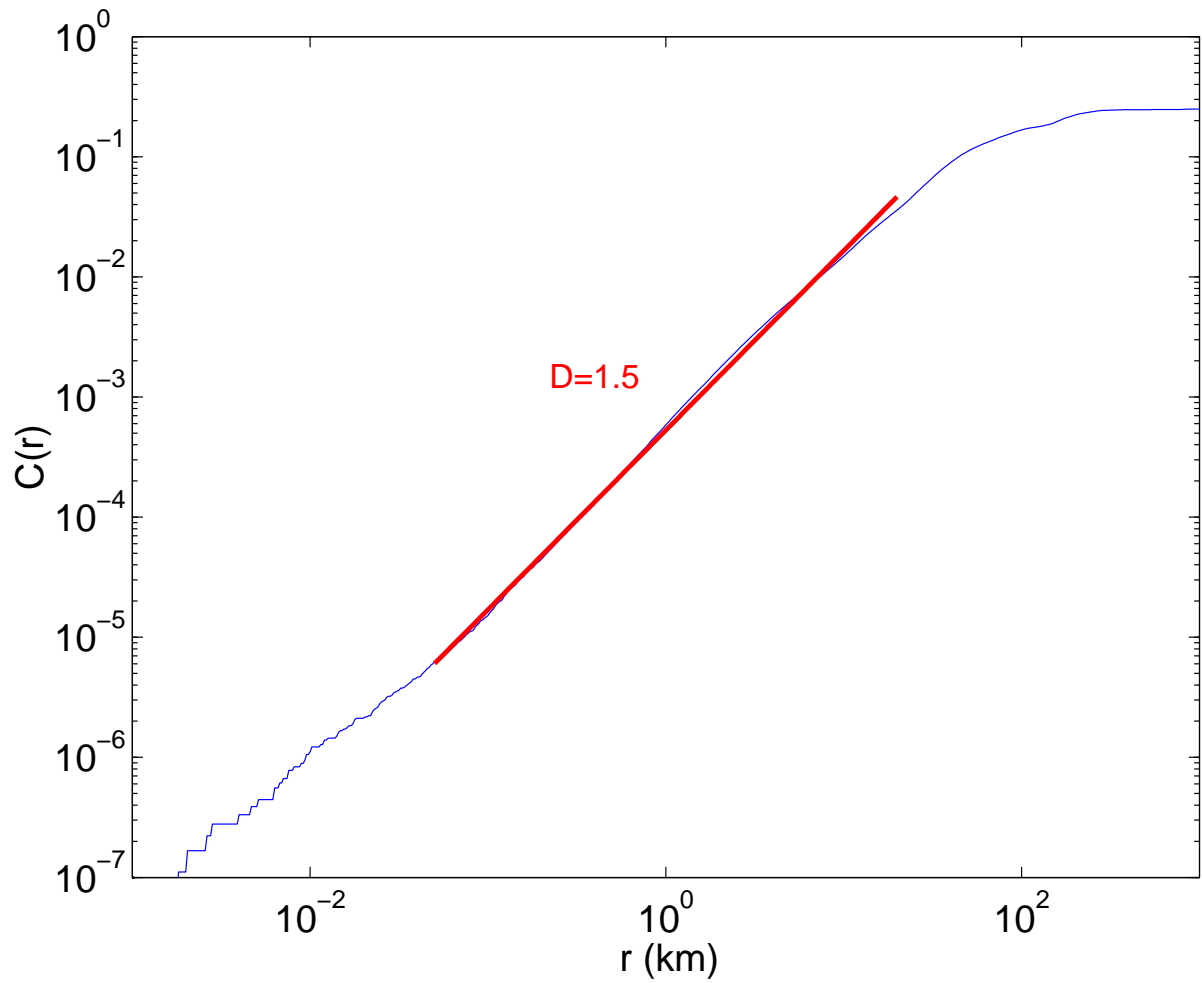}
\caption{\label{cordim} Plot of the correlation function of the $3.000$
epicenters generated in the time interval $[30, 70]$ yrs
and shown in the right panel of figure \ref{map}, calculated
following Grassberger-Procaccia's algorithm \cite{Grassproca}, 
as a function of scale $r$, in double-logarithmic scales.}
\end{figure}

It is interesting to note that the spatial distribution of epicenters shown in
the right panel of figure \ref{map}
has the visual appearance of a fractal set of points.
This is confirmed by the calculation
of the correlation dimension of this set of $N=3000$ points
generated in the time interval $[30, 70]$ yrs, which is found
approximately equal to $D_2=1.5 \pm 0.05$ over more than
two decades in spatial scales, as shown in figure \ref{cordim}.
If we use instead all $30,000$ events of the simulation performed
up to time $t=70$ yrs, we find
$D_2=1.85 \pm 0.05$ while the correlation dimension of the geometrical set
made of the epicenters of the $10,000$ last events (time interval $[7, 70]$ yrs) is
$D=1.7 \pm 0.05$, also over more than two decades in scale. These values
are similar to those reported for 2D maps of active fault systems
\cite{ScholzMandel,Anneexp,BartonLapointe1,Ouillonetal},
and are in good agreement with $D_2$ values in the range $[1.65, 1.95]$
measured for aftershocks epicenters \cite{Nanjo}.
The fractal clustering of the earthquake epicenters,
according to the ETAS model, occurs because of a self-similar process
taking place on many different scales. However, the description
of this multi-scale process solely in terms of a single fractal dimension
fails to fully embody the complex spatial superposition of
local ``singularities'' associated with each aftershock on the one hand
and finite-size effects (stemming from the finite lifetime
of each aftershock sequence) on the other hand. Each event indeed creates
its cloud
of direct aftershocks which can be characterized by its singular exponent
$1-\mu$ for $\mu \leq 1$ and $0$ for $\mu>1$, defined by the scaling 
$\propto\int_0^R r dr/ r^{1+\mu} \propto R^{1-\mu}$ of the ``mass'' 
of the cloud  with its radius $R$. Finite-size effects and 
randomness have been documented to generate realistic but 
sometimes spurious fractal signatures
\cite{Ouillonsor,Hamburger,Eneva,Malcai}. This problem requires
a special study which is left for another work.

\subsection{Relationship with the space-independent ETAS model \label{consosk}}

The spatial ETAS model reduces to the space-independent ETAS model solved
in \cite{HS1} by integrating the dressed propagator
obtained below over all space.
In the Fourier representation (see expression (\ref{ngnslqw})), this
corresponds to putting the wavenumber $k$ to zero. Indeed,
for $k=0$, the Fourier transform amounts to perform a simple integration over
all space. Since ${\hat \Phi}(\vec k=\vec 0)=1$, expression (\ref{ngnslqw})
derived below reduces to the form studied at length in \cite{HS1}.
Therefore, all results reported previously hold also for the version of the
space-dependent ETAS model studied here, when averaging over the whole space.
This is an important property that all the solutions discussed below must obey.

\section{Mapping of the ETAS model on the CTRW model}

In order to study the space-time properties of the ETAS model,
it is very useful to use an exact correspondence between
the ETAS model and the continuous time random walk (CTRW) that we
establish here. In this way,
we can adapt and use the wealth of results previously derived
for the CTRW. But first, let us demonstrate the correspondence
between the ETAS and CTRW models. For this, our
strategy is to derive the Master equation for both models and show
that they are identical. 

\subsection{The Master equation of the ETAS model}

The ETAS model can be rephrased by defining the rate
$\phi_{m_i \to m}(t-t_i, \vec r-\vec r_i)$ at which
a given event (the ``mother'') of magnitude
$m_i \geq m_0$ occurring at time $t_i$ and position $\vec r_i$
gives birth to other events (``daughters'') of specified magnitude
$m$ at a later time
between $t$ and $t+dt$ and at point $\vec r$ to within an
infinitesimal volume $|d{\vec r}|$.
Note that the only difference with respect to the previous definition
(\ref{first})
is that we now specify also the magnitude $m$ of the daughter.
$\phi_{m_i \to m}(t-t_i, \vec r-\vec r_i)$ is given by
\be
\phi_{m_i \to m}(t-t_i, \vec r-\vec r_i) = \rho(m_i \to
m)~\Psi(t-t_i)~\Phi(\vec r-\vec r_i)~,
\label{firstprime}
\ee
where $\Psi(t-t_i)$ and $\Phi(\vec r-\vec r_i)$ are the same as
previously while
\be
\rho(m_i \to m) = P(m)~ \rho(m_i)~.
\ee
With the parameterization (\ref{formrho}) and (\ref{gojfwo}), this reads
\be
\rho(m_i \to m) = n~ \ln(10)~(b - \alpha) ~10^{\alpha
(m_i-m_0)}~10^{-b (m-m_0)}~.
\label{formrhafdgo}
\ee

Let us consider the case where there is an
origin of time $t=0$ at which we start recording the rate of
earthquakes, assuming that
a large earthquake has just occurred at $t=0$ and somehow
reset the clock. In the
following calculation, we will forget about the effect of events
at times prior to $t=0$
and count all aftershocks that are created only by this main shock.

Let us call $N_m(t,\vec r) dt ~dm ~d\vec r$ the number of earthquakes
occurring between
$t$ and $t+dt$ of magnitude
between $m$ and $m+dm$ inside of box of volume $|d{\vec r}|$ centered
at point $\vec r$.
$N_m(t,\vec r)$ is the
solution of a self-consistency equation that formalizes
mathematically the following
process\,: an earthquake may trigger aftershocks; these aftershocks may trigger
their own aftershocks, and so on. The rate of seismicity at a given
time $t$ and position $\vec r$ is the
result of this cascade process. The self-consistency equation that sums up this
cascade reads
\be
N_m(t,\vec r) =  S(t, \vec r, m) +\int {\vec dr}' \int_{m_0}^{\infty} dm'
\int_0^t d\tau~ \phi_{m' \to m}(t-\tau,\vec r-{\vec r}') ~N_{m'}(\tau,
{\vec r}') ~.
\label{third}
\ee
The rate $N_m(t,\vec r)$ at time $t$ and position $\vec r$ is the sum
over all induced rates from all earthquakes of all possible magnitudes
that occurred at all previous times and locations propagated to the present
time $t$ and to the position $\vec r$ of observation by the corresponding
bare propagator. The induced rate
of events per earthquake that occurred at an earlier time $\tau$ and
position ${\vec r'}$ is equal to $\phi_{m' \to m}(t-\tau,\vec r-{\vec r}')$.
The source term $S(t, \vec r)$ is the main shock plus the background
seismicity, if any.
In absence of background seismicity, a main earthquake which occurs
at the origin
of time $t=0$ at position $\vec r= \vec 0$ with magnitude $M$ gives
\be
S(t,\vec r, m)=\delta(t)~\delta(m-M)~\delta(\vec r)
\label{S(t,r,m)}
\ee
where $\delta$ is the Dirac distribution.
Other arbitrary source functions can be chosen.

The source term corresponding to a single mainshock is
indeed the delta function (\ref{S(t,r,m)}) rather than 
the direct Omori law
created by this mainshock in direct lineage. To see this,
notice that the direct Omori law
is recovered from (\ref{third}) by replacing $~N_{m'}(\tau, {\vec r}')$ 
in the integral by $S(t, \vec r, m)$ given by (\ref{S(t,r,m)}). This
shows that the difference between the renormalized 
and the direct Omori laws comes from taking into account the
secondary, tertiary, etc., cascade of aftershocks.

As we have seen, a key assumption of the ETAS model is that the
daughters born from a
given mother have their magnitude drawn independently of the magnitude
of the mother and of the process that give them birth, with a
probability given by the
Gutenberg-Richter distribution (\ref{gojfwo}). The consequences
resulting from relaxing this hypothesis will be reported elsewhere.
Keeping this assumption, it can be shown \cite{HSG} that for $\alpha \leq b/2$
an ensemble of realizations will obey
\be
N_m(t,\vec r) = P(m) ~N(t,\vec r)~,~~~~{\rm for}~~t>0~,  \label{ggnnlalaq}
\ee
which makes explicit the separation of the magnitude from the time
and space variables.
$N(t,\vec r)$ is the number of events at position $\vec r$ at time $t$ of any
possible magnitude. Expression (\ref{ggnnlalaq}) means that the
Gutenberg-Richter
distribution is preserved at all times.
That (\ref{ggnnlalaq}) holds for the ETAS model stems from
the fact that
the waiting time $\Psi(t)$ distribution (\ref{psidef}) and jump size
$\Phi(\vec r)$
distribution (\ref{phidef}) are independent of the magnitudes and that
fluctuations in the seismicity rate are not too wild for $\alpha \geq b/2$.
Note that, in a more complex model in which time, space and magnitudes are
interdependent, expression (\ref{ggnnlalaq}) would become a
mean-field approximation,
in which the fluctuations of the rates induced by the fluctuations of the
realized magnitudes of the daughters factorize from the process.

Putting (\ref{ggnnlalaq}) in (\ref{third}), for $t>0$ when the source
term $S(t,\vec r, m)$ is identically zero, one can simplify by $P(m)$
and obtain
\be
N(t,\vec r) =  \int d\vec r'
\int_0^t d\tau~ \phi(t-\tau,\vec r-\vec r') ~N(\tau, \vec r') ~, ~~~~t>0~,
\label{thirdterter}
\ee
where
\be
\phi(t-\tau,\vec r-\vec r') = \int_{m_0}^{\infty} dm' ~ P(m')
\phi_{m'}(t-\tau,\vec r-\vec r')~.
\label{fngnflq}
\ee

Equation (\ref{thirdterter}) is nothing but the expectation
(or statistical average, i.e., average over an ensemble of
realizations) of expression (\ref{mgmkr}),
with the definition $N(t,\vec r) \equiv {\rm E}[\lambda(t)~\Phi(\vec r)]$.
Therefore, the Master
equation obtained here gives us only the first moment of the space-time
dynamics of seismicity. It is not difficult to derive the equations for the
variance and covariance of the seismic rate as well as higher moments.

The value of the source term at $t=0$ that should be incorporated
in (\ref{thirdterter}) requires more care. Indeed, a naive treatment would give
a source term $\delta(t) \delta(m-M) \delta({\vec r}) / P(M)$ obtained by
simply dividing by $P(m)$, expressed at $m=M$ due to the Dirac
distribution $\delta(m-M)$.
However, this source term still depends on $m$ via the Dirac
distribution $\delta(m-M)$ and
is thus unsuitable as a source term of the equation
(\ref{thirdterter}) which is
independent of $m$. In order to circumvent this difficulty, one has to get
rid of the Dirac distribution $\delta(m-M)$. The corresponding
procedure has been
described in details in Ref.~\cite{HS1} and consists in applying the
integral operator $\int_{m_0}^{\infty} dm ~ {\hat \phi}(\beta,\vec r)$
to (\ref{third}), where ${\hat \phi}(\beta,\vec r)$ is the Laplace 
transform with respect to the time variable of $\phi(t, \vec r)$.
In this way, the Dirac distribution $\delta(m-M)$ is regularized.
Identifying with the results of Ref.~\cite{HS1}, we obtain that
$N(t,\vec r)$ is
solution of (\ref{thirdterter}) with a source term
\be
      S_M(t, \vec r)=\delta(r) \delta(t) \rho(M) /n   ~,
       \label{fngnfldq}
\ee
where $\rho(M)$ is defined in (\ref{formrho}) and $n$ is given by
(\ref{second}).
Thus, the complete Master equation for the
number $N(t,\vec r)$ of events at position $\vec r$ at time $t$ of any
possible magnitude is solution of
\be
N(t,\vec r) =  S_M(t, \vec r) + \int d\vec r'
\int_0^t d\tau~ \phi(t-\tau,\vec r-\vec r') ~N(\tau, \vec r') ~, ~~~~t>0~,
\label{thirdter}
\ee
$N(t,\vec r)$ is the ``dressed'' or ``renormalized'' propagator, obtained
by summing the bare Omori propagator over all possible aftershock cascades.
$N(t,\vec r)$ can also be called the renormalized Omori law \cite{SS}.

The essential assumption used to derive (\ref{third}) is that
the fluctuations of the earthquake magnitudes in a given sequence
can be considered to be decoupled from those of the seismic rate.
This approximation can be shown to be
valid for $\alpha \leq b/2$ \cite{HSG}, for which the random variable $\rho(m_i)$ has
a finite variance.
In this case, any coupling between the fluctuations of the earthquake energies
and the instantaneous seismic rate provides only sub-dominant corrections to the
equation (\ref{third}). For $\alpha > b/2$, the variance of $\rho(m_i)$ is
mathematically infinite or undefined as $\rho(m_i)$ is distributed according
to a power law with exponent $b/\alpha <2$.
In this case, the Master equation  (\ref{third}) is not completely correct
as an additional term must be included to account for the effect of
the dependence between the fluctuations of earthquake magnitudes and
the instantaneous seismic rate. Our results 
are presented below for $\alpha = 0.5$ which belongs to the first
regime $\alpha \leq b/2$. For $\alpha > b/2$, Ref.~\cite{HSG} has shown
that the renormalization of the bare propagator into the dressed propagator
is weaker than for $\alpha \leq b/2$, all the more so as $\alpha \to b$.
Preliminary numerical simulations for $\alpha > b/2$ shows that our results
presented below hold qualitatively but with a reduction of the observed
spatial diffusion exponent compared to the value predicted from the Master
equation approach developed here. This regime $\alpha > b/2$ is probably
relevant to the real seismicity \cite{YaShi,Guo2,H02}, even if a precise
estimation of $\alpha$ is very difficult.

\subsection{A Master equation of the CTRW model}

We now demonstrate that the
self-consistent mean field equation (\ref{thirdter}) is identical to
the Master equation of a
continuous-time random walk (CTRW). Random walks underlie many physical
processes and are often the basis of first-order description of
natural processes.
The CTRW model, which is a generalization of the naive model of a random walker
which jumps by $\pm 1$ spatial step on a discrete lattice at each
time step, was introduced
by \cite{Montroll1} and investigated by many other workers
\cite{Montroll12,Scher,Kenkre,Shlesinger1,Weiss1}.
The CTRW  considers a continuous
distribution of spatial steps as well as time steps (which can be seen
either as waiting times between steps or as durations of the steps).
The CTRW model is thus based on the idea that the length of a given 
jump, as well as the waiting time $\tau_i = t_i - t_{i-1}$
elapsing between two successive jumps are drawn from a joint
probability density function (pdf) $\phi(\vec r, t)$, which is
usually referred to as the jump pdf. From a mathematical point of view, 
a CTRW is a process subordinated to
random walks under the operational time defined by the process $\{t_i\}$.

    From $\phi(\vec r, t)$, the jump length pdf 
$\Phi(\vec r) = \int_0^{+\infty} dt~\phi(\vec r, t)$ and the waiting time 
pdf $\Psi(t) = \int d\vec r ~\phi(\vec r, t)$
can be deduced. Thus, $\Phi(\vec r) d\vec r$ produces the probability
for a jump length in the interval $(\vec r, \vec r+d\vec r)$
and $\Psi(t) dt$ the probability for a waiting time in the interval
$(t, t+dt)$.
When the jump length and
waiting time are independent random variables, this corresponds
to the decoupled form $\phi(\vec r, t) = \Psi(t) ~ \Phi(\vec r)$.
If both are coupled, a jump of
a certain length involves a time cost or, vice versa in a given time
span the walker can only
travel a maximum distance. With these definitions, a CTRW process can
be described
through a Master equation (see \cite{Weiss1,Hughes,Meltzner2}
for a review and references therein) which turns out to be given by an
equation which is identical to (\ref{thirdter}).

This connection between the ETAS model of earthquakes and a model of
random walks provides an important advance for the understanding
of spatio-temporal earthquake processes, as it allows one to borrow
for the deep knowledge accumulated in past decades on random walks.
In the same spirit,
polymer physics acquired its status as a fundamental physical problem
from its previous status of an applied field of research in chemistry
when Flory, Edwards, de Gennes, des Cloizeaux and others showed how
to formulate problems
in polymer physics in the language of random walks and how to extract
novel results. In the sequel of this article, we use this analogy
to provide a wealth of new predictions as well as new questions for
earthquake aftershocks.

In the context of the CTRW, we have the following correspondence.
\begin{itemize}
\item $N(t,\vec r)$ is the pdf for the random walker to just
arrive at position $\vec r$ at time $t$.

\item The source term $S_M(t, \vec r)$ given by (\ref{fngnfldq})
denotes the initial condition of the random walk, here chosen to be
at the origin of space at time $t=0$. The constant $\rho(M) /n$ adds
the possibility via the parameter $M$ to have more than one initial
walker at the origin.

\item In the CTRW context, the Master
equation (\ref{thirdter}) states that the pdf $N(t,\vec r)$ of just
having arrived at position $\vec r$ at time $t$ comes from all possible paths
in number $N(\tau,\vec r')$
having crossed a position $\vec r'$ at an earlier time $\tau$, weighted
by a transfer or propagator function $\phi(t-\tau,\vec r-\vec r')$
describing all the possible steps of the random walker from $(\tau,\vec r')$ to
$(t,\vec r)$.
\end{itemize}

It is important to stress that $N(t,\vec r)$ defined above
is different from the standard quantity $W(t,\vec r)$ usually studied
in random walk problems,
defined as the probability to find the random walk at position $\vec
r$ at time $t$.
The relationship between $N(t,\vec r)$ and $W(t,\vec r)$ is
\be
W(t,\vec r) = \int_0^t dt'~\left[ 1 - \int_0^{t-t'} dt''~
\Psi(t'')\right]~N(t',\vec r)~.
\label{bheiw}
\ee
The term $1 - \int_0^{t-t'} dt''~ \Psi(t'')$ in bracket is the probability for
the walker not to jump in the time interval $[t',t]$ and the integral
in the right-hand-side of (\ref{bheiw}) means that the probability
$W(t,\vec r)$ for the random
walker to be at position $\vec r$ at time $t$ is the sum over all possible
scenarios in which the walker just arrives at $\vec r$ at an earlier
time $t'$ and
then does not jump until time $t$. In the context of earthquake aftershocks,
$W(t,\vec r)$ is the probability that
an event at $\vec r$ has occurred at a time $t' \leq t$
and that the whole system has remained quiescent from $t'$ to $t$.

In the Fourier-Laplace domain (see below), expression (\ref{bheiw}) reads
\be
{\hat W}(\beta,\vec k) = {1 - {\hat \Psi}(\beta) \over \beta}~{\hat
N}(\beta,\vec k)~.
\label{mhgjhjhd}
\ee

In general, the CTRW models transport phenomena in any
heterogeneous media. It has for instance been used successfully for
describing the behavior
of chemical species as they migrate through porous media
\cite{Margolin,Berko1}.
In insight, it is rather natural that it can be applied to the
``transport of stress''
through the heterogeneous crust and thus to the description of the anomalous
diffusion of seismic activity.

Table \ref{table1} synthesizes
the correspondence between the ETAS and CTRW models and
then draws its consequences.

\subsection{Experimental verifications of the cross-over between the
two power law Omori decays in photoconductivity in amorphous semi-conductors
and in fractal stream chemistry
using the correspondence between the ETAS and CTRW model}

The crossover from an Omori law
$1/t^{1-\theta}$ for $t<t^*$ to
$1/t^{1+\theta}$ for $t>t^*$ found in
\cite{SS,HS1} with $t^*$ given by (\ref{tstar})
has actually a counterpart in the CTRW. This
behavior was first studied by Scher and Montroll \cite{Scher} in a CTRW with
absorbing boundary condition to model photoconductivity in amorphous
semi-conductors
As$_2$Se$_3$ and an organic compound TNF-PVK finding $\theta \approx 0.5$ and
$\theta = 0.8$ respectively. In a semiconductor experiment, electric holes
are injected near a positive electrode and then transported to a negative
electrode where they are absorbed. The transient current follows exactly
the transition $1/t^{1-\theta}$ for $t<t^*$ to
$1/t^{1+\theta}$ for $t>t^*$ found for Omori law for earthquake aftershocks
in the ETAS model. In the semiconductor context, the finiteness of $t^*$
results from the existence of a force applied to the holes while in the ETAS
model it results from a finite distance $1-n$ to the critical point $n=1$
in the subcritical regime. When the force goes to zero or $n \to 1$,
$t^* \to +\infty$.

A similar transition has been  recently proposed to model
long-term time series measurements of chloride, a natural passive tracer,
in rainfall and runoff in catchments \cite{Schermarklaber}. The
quantity analogous to
the dressed Omori propagator is the effective travel time distribution $h(t)$
which governs the global lag time between injection of the tracer
through rainfall
and outflow to the stream. $h(t)$ has been shown to have a power-law form
$h(t) \sim 1/t^{1-m}$  with $m$ between -0.3 and 0.2 for different time series
\cite{Kirchner}. This variability may be due to the transition
between an exponent
   $1-\theta$ at short times to $1+\theta$ at long times \cite{Schermarklaber},
where $\theta$ is the exponent of the bare distribution of individual
transition times.

\subsection{General and formal solution of the spatial ETAS model}

Let us solve (\ref{thirdter}) for the
number $N(t,\vec r)$ of events at position $\vec r$ at time $t$ of any
possible magnitude. Recall that $N(t,\vec r)$ can also be interpreted
as the dressed Omori propagator. Extending \cite{HS1}
to the spatial domain and also in analogy with the standard approach to
solve the CTRW, the Laplace-in-time Fourier-in-space transform ${\hat
N}(\beta,\vec k)$ of $N(t,\vec r)$ is given by
\be
{\hat N}(\beta,\vec k) = {{\hat S}_M(\beta,\vec k) \over
1 - n {\hat \Psi}(\beta) {\hat \Phi}(\vec k)}~,
\label{ngnslqw}
\ee
where ${\hat S}_M(\beta,\vec k)$ is the Laplace Fourier transform of the source
$S_M(t, \vec r)$ given by (\ref{fngnfldq})
and ${\hat \Psi}(\beta)$ (respectively ${\hat \Phi}(\vec k)$) is the Laplace
(respectively Fourier) transforms of $\Psi(t)$ (respectively $\Phi(\vec r)$).
For a mainshock of magnitude $M$ occurring at time $t=0$ and position
$\vec r=0$,
the source term is thus ${\hat S}_M(\beta,\vec k)=\rho(M)/n$.
The only difference between expression (\ref{ngnslqw}) and the Laplace-Fourier
transform of the pdf of the CTRW of just having arrived at $\vec r$ at time $t$
occurs when the branching ratio $n$ is different from $1$.  In general,
solutions of CTRW models are expressed for $n=1$ and for the variable
$W(t, \vec r)$
which is simply related to $N(t, \vec r)$ according to (\ref{bheiw}). Using
(\ref{bheiw}) and (\ref{ngnslqw}) leads to
\be
{\hat W}(\beta,\vec k) = {1 - {\hat \Psi}(\beta) \over \beta}~~{{\hat
S}_M(\beta,\vec k) \over
1 - n {\hat \Psi}(\beta) {\hat \Phi}(\vec k)}~,
\label{ngnslaaqw}
\ee

In the following, we exploit (\ref{ngnslaaqw}) to
obtain analytical solutions of the spatial ETAS model in
different regimes,
that provide specific predictions on the conditions necessary for observing
aftershock diffusion. In addition, we provide specific predictions
on the exponent $H$ of the
diffusion law $R \sim t^{H}$ that are tested by numerical simulations.

\section{Critical regime $n=1$}

\subsection{Classification of the different regimes}

Numerous works on the CTRW have investigated many possible forms for
$\Psi(t)$ and $\Phi(\vec r)$ and have provided the asymptotic long time
and large scale dependence of $W(t, \vec r)$ (see
\cite{Weiss1,Hughes,Meltzner2,Berko1} and references therein).
Here, we restrict our discussion to the cases where both $\Psi(t)$ and
$\Phi(\vec r)$ have power law tails as given by (\ref{psidef}) and
(\ref{phidef}).
The long-time and large scale behavior of the ETAS and CTRW are controlled by
the behavior of the Laplace-Fourier transforms for small $\beta$ and
small $|\vec k|$.

Two cases must be distinguished depending on the exponent $\mu$ controlling
the weight of the tail of $\Phi(\vec r)$.
\begin{itemize}
\item For $\mu > 2$, the variance $\langle (\vec r)^2 \rangle =
\sigma^2$ of the jump
size distribution exists. To leading order in $k = |\vec k|$, ${\hat
\Phi}(\vec k)$
can be expanded as
\ba
{\hat \Phi}(\vec k) = 1- \sigma^2 k^2 + {\cal O}(k^o)~,~~~{\rm with}~~o>2~.
\label{ngnglw}
\ea

\item For $\mu\leq 2$, the variance $\langle (\vec r)^2 \rangle$
is infinite. This regime of ``long jumps'' leads to so-called L\'evy flights.
In this case, to leading order in $k = |\vec k|$, ${\hat \Phi}(\vec k)$
can be expanded as
\be
{\hat \Phi}(\vec k) = 1- \sigma^{\mu} k^{\mu} + {\cal O}(k^o)~,~~~{\rm where}~~
0 < \mu \leq 2, ~~~{\rm with}~~o>\mu~,
\label{ngngfdlw}
\ee
where $\sigma$ is a characteristic distance defined by
\be
       \sigma= \left\{ \begin{array}{lll}
       d~[\Gamma(1-\mu)]^{1/\mu}    , & &  0<\mu < 1~, \\
       {d~\pi \over \mu ~ \Gamma(\mu-1)~\sin(\pi \mu /2) }  , & & 1<\mu<2~.
       \end{array}
       \right.
       \label{fklj}
\ee
\end{itemize}

For a distribution $\Psi(t)$ of waiting times of the form
of a local Omori law (\ref{psidef}) with exponent $\theta < 1$,
${\hat \Psi}(\beta)$ can be expanded for small $\beta$ as
\be
{\hat \Psi}(\beta) = 1 - (\beta c')^{\theta} + {\cal
O}(\beta^\omega)~,~~~{\rm with}~
\omega \geq 1~.
\label{hgngwlff}
\ee
where $c'$ is proportional to $c$ up to a numerical constant
$c'=c \left(\Gamma(1-\theta)\right)^{1/\theta}$ in the case $\theta<1$.

Putting the leading terms
of the expansions of ${\hat \Phi}(\vec k)$ for small $|\vec k|$ and
of ${\hat \Psi}(\beta)$
for small $\beta$ in (\ref{ngnslqw}) gives
\be
{\hat N}(\beta,\vec k) = {{\hat S}_M(\beta,\vec k) \over 1-n +
n(\beta c')^{\theta} + n\sigma^{\mu} k^{\mu}}~.
\label{ngnslqwassa}
\ee
The corresponding ${\hat W}(\beta,\vec k)$ is obtained from
(\ref{ngnslaaqw}) by
\be
{\hat W}(\beta,\vec k) = {\hat S}_M(\beta,\vec k)~{(\beta)^{\theta
-1}  c'^{\theta} \over
1-n +n (\beta c')^{\theta} + n\sigma^{\mu} k^{\mu}}~.
\label{ngnslqssaawaa}
\ee

The critical regime $n=1$ gets rid of the constant term $1-n$ in the
denominator of (\ref{ngnslqwassa}) and (\ref{ngnslqssaawaa}). This case
is analyzed in details below.

The regime $n\neq 1$ introduces a characteristic time $t^*$ given by
(\ref{tstar}).
In the sub-critical regime, equation (\ref{ngnslqwassa}) can be rewritten as
   \be
{\hat N}(\beta,\vec k) = {{\hat S}_M(\beta,\vec k) \over (1-n)}~
{1 \over 1 + (\beta t^*)^{\theta} + (k r^* )^{\mu}}~.
\label{ngnslqwassa2}
\ee
where $r^*$ is defined by
\be
r^*=\sigma \left({n \over 1-n}\right)^{1/\mu}~.
\label{gnjgrkd}
\ee
For $t<t^*$ and $r<r^*$, the dressed propagator is given by the same
expression as
for the critical case and all our results below hold.
For large times  $t>t^*$ and large distances $r>r^*$,
we can factorize   (\ref{ngnslqwassa2}) as a
product of a function of time and a function of space
   \be
{\hat N}(\beta,\vec k) \simeq {{\hat S}_M(\beta,\vec k) \over (1-n)}
{1 \over (1+(\beta t^*)^{\theta})}{1 \over (1 + (k r^*)^{\mu})}~.
\label{ngnslqwassa3}
\ee
Thus, there is no diffusion in the sub-critical regime for $t>t^*$ and $r>r^*$.
We shall not analyze further this
trivial regime $n<1$ and $t>t^*$ and will only analyze the case $t<t^*$.
If there is the need, the cross-over can be calculated
explicitly using (\ref{ngnslqwassa}).

In order to get the leading behavior of $N(t, \vec r)$ from that of
$W(t, \vec r)$,
we see from (\ref{ngnslqw}) and (\ref{ngnslaaqw}) that
${\hat N}(\beta,\vec k)  = {\beta \over 1 - {\hat \Psi}(\beta)}~{\hat
W}(\beta,\vec k)
\approx \beta^{1-\theta} c'^{-\theta} ~{\hat W}(\beta,\vec k)$. The
inverse Laplace transform
of $1/\beta^{\theta}$ is $1/[\Gamma(\theta)~t^{1-\theta}]$. Using the
fact that the Laplace transform of $df/dt$ is $\beta$ times the
Laplace transform
of $f(t)$ minus $f(0)$, we get $N(t, \vec r)$ as the
derivative of a convolution
\be
N(t, \vec r) = {c'^{-\theta} \over \Gamma(\theta)}
{d \over dt}~\int_0^t dt'~ {W(t', \vec r) \over (t-t')^{1-\theta}}
= c'^{-\theta}~ _0D_t^{1-\theta}~ W(t, \vec r)~.
\label{nbjn}
\ee
In (\ref{nbjn}), we have dropped the Dirac function coming from the inverse
Laplace transform of the constant term $f(0)$, which provides
a contribution only at the origin of time $t=0$. Note that the
operator ${1 \over \Gamma(\theta)}
{d \over dt}~\int_0^t dt'~ {W(t', \vec r) \over (t-t')^{1-\theta}}$
is nothing but the so-called fractional Riemann-Liouville derivative operator
of order $1-\theta$ applied to the function $W(t, \vec r)$ of time $t$ and is
    usually denoted
$_0D^{1-\theta}_t W(t, \vec r)$.

\subsection{The standard diffusion case $\theta >1$ and $\mu>2$ \label{mgmls}}

The standard diffusion process is recovered for $\theta \geq 1$
(for which the average waiting time is finite)
and for $\mu \geq 2$ (for which the variance of the jump length is finite).
In this case,
${\hat N}(\beta,\vec k) = {{\hat S}_M(\beta,\vec k) \over
\beta c' + \sigma^{2} k^{2}}$. For an impulsive source
leading to ${\hat S}_M(\beta,\vec k) =$ constant, this is the
Laplace-Fourier transform of the standard diffusion propagator
\be
N(t, \vec r) \propto {1 \over (D t)^{d/2}}~
\exp [ - (\vec r)^2/Dt]~,~~~~{\rm where}~~D = \sigma^{2}/c'~,
\label{gngk}
\ee
where $d$ is here the space dimension.
This solution is valid for $|\vec r|/\sqrt{Dt}$ not too large. For
larger values,
large deviations lead to corrections with the power law tail of the
input jump distribution $\Phi(\vec r) \sim 1/|\vec r|^{1+\mu}$ defined in
(\ref{phidef}), along the lines presented for instance in
\cite{Sorbook} (section 3.5). This regime is not relevant to the
aftershock problem for which usually $0 < \theta <1$.

\subsection{Long waiting times ($\theta < 1$) and finite variance
of the jump sizes ($\mu>2$)}

Putting the leading terms of the expansions of ${\hat \Phi}(\vec k)$
(\ref{ngnglw}) and of ${\hat \Psi}(\beta)$ (\ref{hgngwlff})
in (\ref{ngnslqw}) gives
\be
{\hat N}(\beta,\vec k) = {1 \over (\beta c')^\theta + (\sigma k)^2}
\label{ngbnmvd}
\ee

The expression (\ref{ngbnmvd}) can be inverted with respect to the
Fourier transform, and then
inverted with respect to the Laplace transform using Fox functions
\cite{Meltzner2,barkai3}.
The solution for $W(t, \vec r)$ in one dimension is given for instance
in \cite{Meltzner2} in terms of an infinite sum
\be
W(t,\vec r) ={ 1 \over 2 D}~{1 \over t^{\theta \over 2}}~
\sum_{k=0}^{\infty} {(-1)^k ~ z^{-k} \over k!~\Gamma(1-\theta (k+1)/2)}
\label{giouyt}
\ee
where
\be
z= { D~t^{\theta/2} \over  |\vec r|}
\label{zdef}
\ee
and $D=\sigma /c'^{\theta/2}$.   

Expression (\ref{giouyt}) and many others below involve the Gamma function of
negative arguments. We recall that the Gamma function $\Gamma(u)$ can be
analytically continued to the whole complex plane, except for the simple poles
$u=0, -1, -2, -3, ...$ Thus, $\Gamma(u)$
is defined everywhere but at these poles.
In order to get the expression of the Gamma
function for negative arguments, one can use two formulae:
$\Gamma(1-u) \times \Gamma(u) = \pi /\sin(\pi u)$ and
$\Gamma(1+u) = u  \Gamma(u)$.
Both these formulae are valid  for all points with the possible exception of
the arguments at poles $0, -1, -2,...$ For instance,
$\Gamma(-\theta) = \Gamma(1-\theta) / (-\theta) =
-[\pi /\theta \sin(\pi \theta)]/ \Gamma(\theta)$, for $0 < \theta < 1$.

Expression (\ref{giouyt}) can be rewritten as a Fox-function \cite{Mathai}
\be
W(t,z) = {1 \over 2D}~ {1 \over t^{\theta \over 2}}~ H_{1,1}^{1,0}
\left[{1 \over z} \left |
{\begin{array}{l} (1-\theta/2,\theta/2) \\ (0,1) \end{array}}
\right .
\right ]
\label{fox4346}
\ee
whose asymptotic dependence for large $z$,
obtained from a standard theorem of the Fox function (equation (1.6.3) of
\cite{Mathai}),
\be
W(t, z) \sim {1 \over D~t^{\theta \over 2}}~
{1 \over z^{1-\theta \over 2-\theta}}~
\exp \left(-\left(1-{\theta \over 2}\right)\left({\theta \over
2}\right)^{\theta \over 2- \theta}
z^{2 \over 2-\theta} \right )
\label{exp56}
\ee
is in agreement with the result of Roman and Alemany \cite{Roman} and
Barkai et al. \cite{barkai3}
for a space dimension $d_f=1$,
including the dependence in the power law prefactor to the exponential.
The exponential dependence $W(t, r) \sim
\exp \left(- {\rm const}~ (r/D t^{\theta/2})^{2 \over 2-\theta} \right )$
in (\ref{exp56}) holds in arbitrary dimensions $d_f$, the only
modification occurring
in the prefactor whose power of $z$ change with the space
dimension $d_f$ as \cite{Roman,barkai3}
\be
W_{d_f}(t, z) \sim {1 \over D~ t^{{\theta \over 2}}}~
{1 \over z^{d_f (1-\theta) \over 2-\theta}}~
\exp \left(-\left(1-{\theta \over 2}\right)\left({\theta \over
2}\right)^{\theta \over 2- \theta}
z^{2 \over 2-\theta} \right )~.
\label{exp5a6}
\ee

The expression of $N(t,\vec r)$ can be obtained from $W(t,\vec r)$ 
using the fractional
Riemann-Liouville derivation (\ref{nbjn}) of order $1-\theta$.
Inserting expression (\ref{giouyt}) in (\ref{nbjn})
and using the expression of
the fractional Riemann-Liouville derivative operator
$_0D^{\alpha}_t$ applied to an arbitrary power $t^{\mu}$, i.e.,
$_0D^{\alpha}_t t^{\mu} = {\Gamma (1+\mu) \over
\Gamma(1+\mu-\alpha)}~t^{\mu-\alpha}$,   we obtain
\be
N(t,\vec r) = {c'^{-\theta} \over 2 D t^{1-{\theta \over 2}}}~
\sum_{k=0}^{\infty} {(-1)^k ~ z^k \over k!~ \Gamma((1-k)\theta /2)}~.
\label{qdsi}
\ee
    Expression (\ref{qdsi}) can be used to evaluate $N(t,\vec r)$ for small $z$,
    but the numerical evaluation of (\ref{qdsi}) is impossible for large $z$.
    In order to obtain the asymptotic behavior of $N(t,\vec r)$,
    expression (\ref{qdsi}) can be rewritten as a Fox-function \cite{Mathai}
\be
N(t,\vec r) = {c'^{-\theta} \over 2D t^{1-{\theta \over 2}}}~ H_{1,1}^{1,0}
\left[ {1 \over z } \left |
{\begin{array}{l} (\theta/2,\theta/2) \\ (0,1) \end{array}}
\right .
\right ]  ~.
\label{fox3}
\ee
Employing again the standard theorem of the Fox function (equation (1.6.3) of
\cite{Mathai}),
the asymptotic behavior of $N(t, r)$  for large distances $r$
such that $r>D t^{\theta/2}$ is given by
\be
N(t, r) \sim {c'^{-\theta}  \over D t^{1-{\theta \over 2}}}~
\left({|\vec r| \over D t^{\theta/2}} \right)^{1-\theta \over 2-\theta}~
\exp \left(-\left(1-{\theta \over 2}\right)\left({\theta \over
2}\right)^{\theta \over 2- \theta}
\left({|\vec r| \over D t^{\theta/2}} \right)^{2 \over 2-\theta} \right )~.
\label{exp1}
\ee
The exponential dependence $N(t, r) \sim
\exp \left(- {\rm const}~ (r/D t^{\theta/2})^{2 \over 2-\theta} \right )$
in (\ref{exp1}) holds in arbitrary dimensions.

This expression becomes incorrect for very large distances because it would
predict an exponential or slightly super-exponential decay with $r$.
This cannot be true as the global law cannot decay faster than the
local law (\ref{phidef}). The reason for (\ref{exp1}) to become incorrect at
large distances is that the expansion of
${\hat N}(\beta,\vec k)$ for small $|\vec k|$ (large distances) given by
(\ref{ngbnmvd}) has been truncated at the order $k^{2}$. There is however
a subdominant term $\propto k^{\mu}$ that describes the power law tail
of the local law (\ref{phidef}) and also of the global law asymptotically.
A similar situation occurs in the application of the
central limit theorem for sums of $N$ random variables with power law
distributions with exponents $\mu>2$ \cite{Sorbook}: the distribution
of the sum $S$
is a Gaussian in its bulk for $|S| < \sqrt{N \ln N}$ and crosses over
to a power law
with tail exponent $\mu$ for larger $S$. In a similar way, the cross-over
of $N(t, r)$ to the
asymptotic local power law (\ref{phidef}) can be recovered by an analysis
including the subleading correction $\propto k^{\mu}$ to the expansion
(\ref{ngbnmvd}).

Expression (\ref{qdsi}) shows that the global rate of seismicity cannot be
factorized as a product of a distribution of times and a distribution
of distances.
This space-time coupling implies that the seismic activity diffuses with time,
and that the decay
of the rate of aftershocks depends on the distance from the first mainshock.
This coupling of space and time stems from the cascade of
aftershocks, from the primary aftershocks to the secondary aftershocks to the
tertiary aftershocks and so on.

\begin{figure}
\includegraphics[width=9cm]{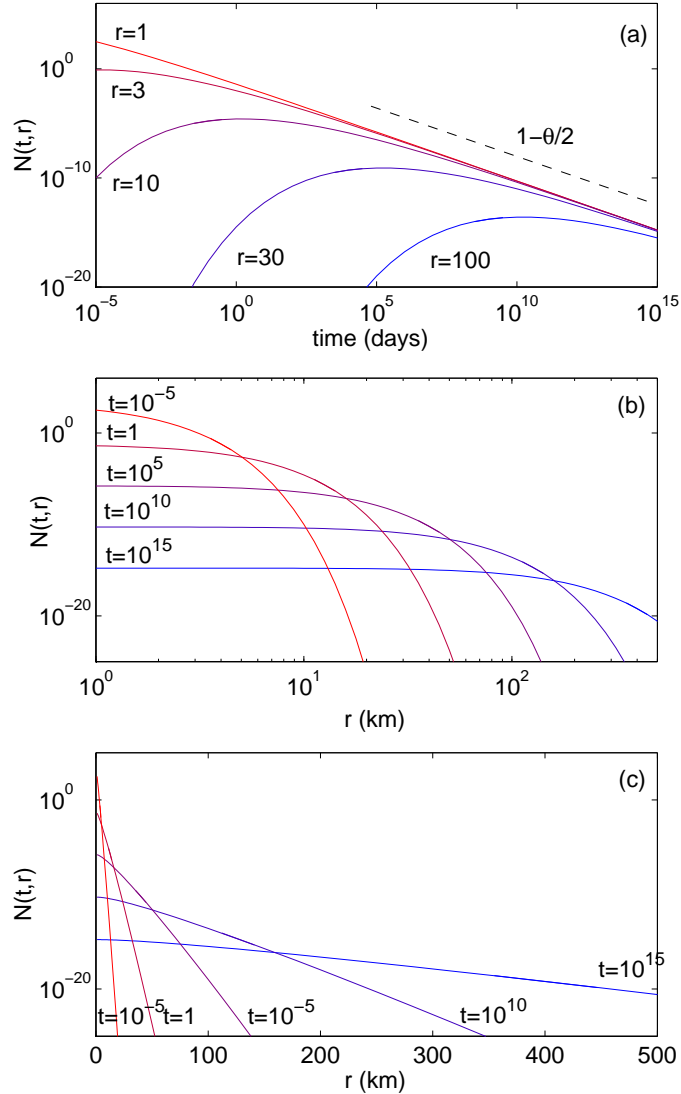}   
\caption{\label{nrtmus2} Rate of seismicity $N(t,r)$ in the critical
regime $n=1$
  for $\theta=0.2$, $\mu>2$, $c'=1$ day and $\sigma=1$ km,
evaluated from expressions (\ref{qdsi}) and (\ref{exp1}), plotted as a function
  of the time (a) for different values of the distance $r$ between the mainshock
   and its aftershocks, and (b,c) as a function of $r$ (logarithmic scale
   for $r$ in (b) and linear scale for $r$ in (c)) for different values of
   the time between the mainshock and its aftershocks.
   The temporal decay of seismicity with time is characterized by a
power-law decay
   $ N(r,t) \sim {1 / t^{1-\theta/2}} $ close to the mainshock epicenter or at
    large times for $r \ll D t^{\theta/2}$.
  For large distances $r \gg D t^{\theta/2}$, there is a truncation of the
  power-law decay at early times $t^{\theta / 2} \ll r/D$, because the
seismicity
   has not yet diffused up to the distance $r$.
   Although the distribution of distances between a mainshock and its
direct aftershocks
   $\Phi(r)$ follows a power-law distribution with exponent $1+\mu$,
the log-linear
   graph (c) shows that the global
   rate of aftershocks $N(\vec r,t)$ decreases approximately
   exponentially as a function of
   the distance from the mainshock, with a characteristic distance
that increases with time.
   This is in agreement with expression (\ref{exp1}) which predicts
   $N(t, r) \sim \exp \left[\left( |\vec r| / D t^{\theta/2}
\right)^{2 \over 2-\theta}\right]$,
   i.e., $N(t, r) \sim \exp \left( C(t) |\vec r|^q \right)$ with an
   exponent $q=2 /(2-\theta)$ close to $1$ within the exponential. The
same remark as for
   figure \ref{n} applies: the representation of our predictions for
very large times
   is made for pedagogical purpose to illustrate clearly the different regimes.
}
\end{figure}

Figure \ref{nrtmus2} presents the decay of the seismic activity
$N(r,t)$ obtained
using expression (\ref{qdsi}) for small $z$ and expression
(\ref{exp1}) for large $z$,
as a function of the time from the mainshock and as a function of the
distances $r$. Close to the mainshock epicenter, expression (\ref{qdsi})
predicts that the global seismicity rate decays with time as the 
renormalized Omori law
\be
N(t,0) \sim  {1 \over t^{1-\theta/2}}~.
\label{oyu}
\ee
The same decay is found at any fixed point $\vec r$ for times $t >
(|\vec r|/D)^{2/\theta}$.  At all times, the same decay $1/t^{1-\theta/2}$
     is also obtained by measuring the
aftershock seismicity in a local box at a distance from the main shock origin
increasing with time as $r \sim t^{\theta \over 2}$ (this is nothing
but putting $z=$ constant in (\ref{qdsi})).
At large distances  $r > D t^{\theta/2}$, the global decay law is
different from a
power-law decay. Figure \ref{nrtmus2} shows that the rate of
aftershocks presents
a truncation at early times, which increases as the distance $r$ increases.
At large times, the rate of
aftershocks recovers the $1/t^{1-\theta/2}$ power-law decay (\ref{oyu}).
We stress that a fit of the global law $N(r,t)$ over the whole
time interval by an Omori law would yield an apparent
exponent  $p< 1-\theta/2$ that decreases with $r$.

Integrating (\ref{qdsi}) over the whole one-dimensional space, we recover the
global Omori law
\be
N(t)=\int dr N(t,r) \sim {1 \over t^{1-\theta}}
\ee
found in \cite{SS,HS1}. Thus, we have found an additional source of variability
of the exponent $p$ of the Omori law: if measured over the whole catalog,
we should measure $p=1-\theta$ in the critical regime $n=1$ while
$p=1-\theta/2$ is
slightly larger when measured in certain time- and space-windows,
as described above. Thus, in this regime,
pruning of catalogs may lead to continuous change
from the value $1-\theta$ to $1-\theta/2$. In addition, as we have mentioned,
the cross-over in time may lead to still smaller apparent exponents, thus
enhancing the impression of variability of the exponent $p$. In reality, this
range of $p$-values are seen to result from the complex spatio-temporal
organization of the aftershock seismicity of the ETAS model. These results
should lead us to be cautious when analyzing real catalogs with respect
to the conditions and regimes under which the analysis is performed.

There is another observable that
characterizes how an aftershock sequence invades space as a function of time.
Expression (\ref{qdsi}) indeed predicts a sub-diffusion process quantified by
\be
\langle |\vec r|^2 \rangle  \sim t^{2H}~, \label{hgnvaa}
\ee
with $H = \theta/2$ since the natural variable is
$z$ given by (\ref{zdef}).
Indeed, expression (\ref{qdsi})
tells us that, up to a global rescaling function of time, the rate of
aftershocks is identical for a fixed value of $z$. Thus, any
aftershock structure
diffuses according to (\ref{hgnvaa}).

\begin{figure}
\includegraphics[width=16cm]{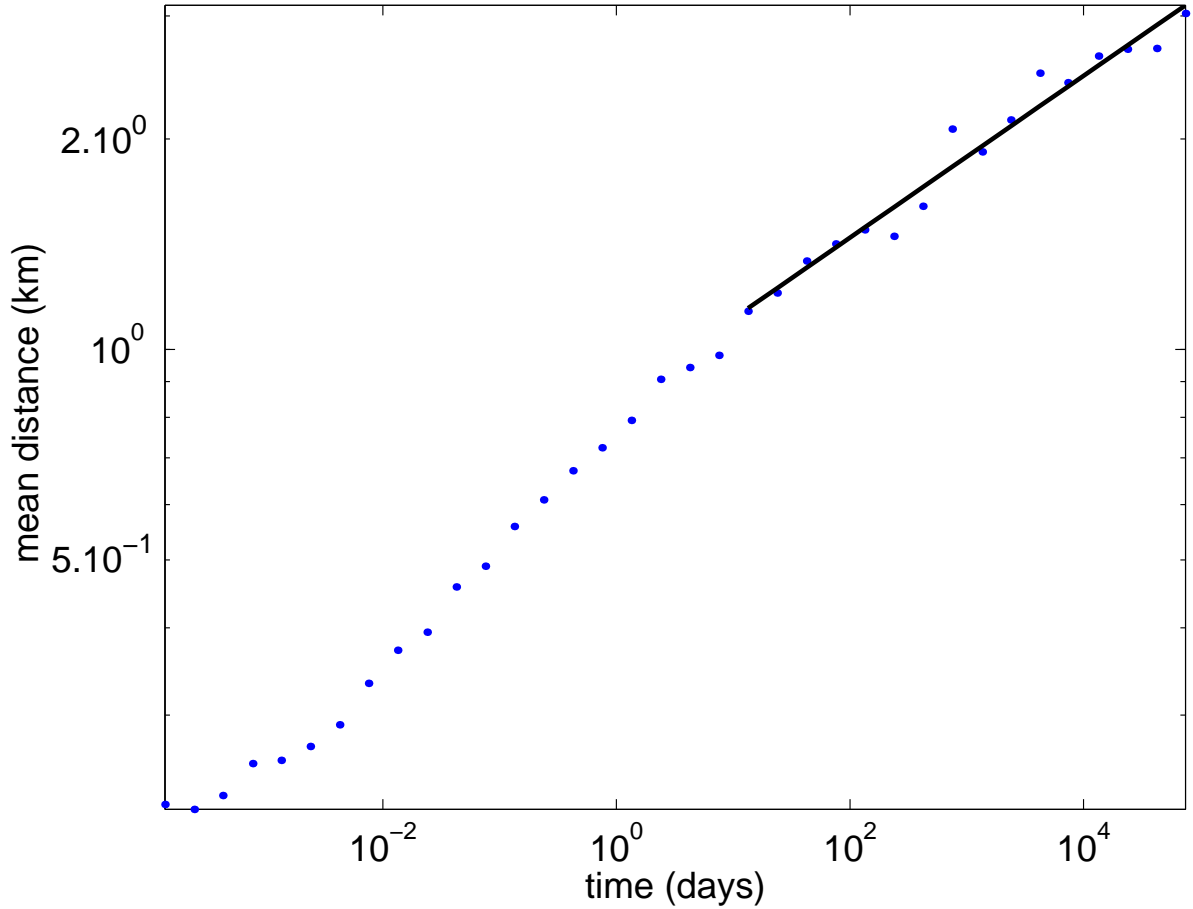}  
\caption{\label{poisq} Average distance between the first mainshock and its
aftershocks as a function of the time from the mainshock, for numerical
simulations of the ETAS model in the critical regime $n=1$, generated with the
   parameters $\theta=0.2$, $d=1$ km, $\mu=3$ and $c=10^{-3}$ day.
The theoretical prediction for the diffusion exponent is thus $H=\theta/2=0.1$.
We observe a crossover from a larger exponent at early times when the
mean distance
    is close to the characteristic scale $d=1$ km of the distribution of
distances between
    an aftershock and its progenitor,
to a sub-diffusion with an exponent close to the theoretical prediction at
large times. The solid line is a fit of the numerical data for times $t>10$ days,
which gives an exponent $H=0.12$ slightly larger than the predicted
value $H=0.1$.
}
\end{figure}

This prediction is checked in Figure \ref{poisq} by numerical simulations.
1000 synthetic catalogs have been generated with $\mu=3$, $\theta=0.2$
and $n=1$. The average distance between the first mainshock
and its aftershocks as a function of the time from the mainshock has
been averaged over these 1000 simulations.
The theoretical diffusion exponent is  $H=\theta/2=0.1$, in
good agreement with the
asymptotic behavior observed in the numerical simulation.
In practice, in order to minimize the effect of fluctuations and optimize
the speed of convergence, we estimate numerically $\exp [\langle \ln
|\vec r| \rangle]$ which is also expected to scale as
$\exp [\langle \ln |\vec r| \rangle] \sim t^{\theta/2}$
due to the simple scaling form of (\ref{fox3}).

This problem has also been solved exactly in \cite{barkai1} in the context
of the so-called fractional Fokker-Planck equation, which amounts to
replace the distribution $\Phi(\vec r)$ of jumps (\ref{phidef}) by a
Gaussian function.
This fractional Fokker-Planck equation allows one to introduce the
possibility of bias
or drift in the CTRW and therefore in the aftershock sequence.

\subsection{Exponential waiting time distribution and long jump size
L\'evy distribution ($\mu<2$)}

This case with an exponential distribution
\be
\Psi(t) = \lambda ~e^{-\lambda t}
\label{ngllsa}
\ee
of waiting times with a L\'evy distribution
$\Phi(\vec r)= L_{\mu}(|\vec r|)$ of jump sizes with tail exponent $\mu<2$
has been investigated  by Budde et al. \cite{Budde}.
One finds
\be
\langle |\vec r|^2 \rangle^{1/2} \sim t^{1/\mu}~, \label{hgnvaaaa}
\ee
corresponding to a superdiffusion regime with Hurst exponent $H=1/\mu >1/2$.
The full distribution function $W(t, \vec r)$ corresponding to the
critical regime $n=1$ is known for $\lambda t >> 1$:
\be
W(t, \vec r) \propto {1 \over (\lambda t)^{1/\mu}}~ L_{\mu}\left({|\vec r|
\over (\lambda t)^{1/\mu}}\right)~.
\label{mgmkrl}
\ee
The corresponding $N(t, \vec r)$ is obtained from (\ref{mhgjhjhd}).
The Laplace transform of the exponential distribution (\ref{ngllsa})
is ${\hat \Psi}(\beta)= \lambda/(\beta + \lambda)$. We thus get
\be
{\hat N}(\beta,\vec k) =
\left( \beta + \lambda \right) ~{\hat W}(\beta,\vec k)~,
\ee
and thus
\be
N(t, \vec r) = {\partial W(t, \vec r) \over \partial t}~+~\lambda
~W(t, \vec r)~.
\label{mgkjhgk}
\ee
Expression (\ref{mgkjhgk}) together with (\ref{mgmkrl}) predicts
a diffusion law $r \sim t^H$ with $H=1/\mu$ which is in good agreement
with our simulations.
At large times $|{\vec r}| \ll (\lambda t)^{1/\mu}$,
$N(t, \vec r) \approx \lambda ~W(t, \vec r) \sim 1/t^{1/\mu}$, given
an apparent local Omori exponent $\theta = 1 - 1/\mu$.
This offers a new mechanism for generating Omori law for aftershocks
from purely
exponential local relaxation but with a heavy distribution of jump sizes.
This power-law decay should be observed only at a fixed distance $r$
or over a limited domain from the mainshock
in the regime of large times.

Integrating over the whole space, $\int d\vec r ~W(t, \vec r) =1$ which gives
$N(t) = \delta(t) + \lambda$ equal to a constant seismic rate. This
results from an initial mainshock at $t=0$ leading to the cascade of
aftershocks
adjusting delicately to this constant rate for the critical value $n=1$ of
the branching parameter. In the sub-critical regime $n<1$, the Omori law
integrated over space gives instead $N(t) \propto \exp [-(1-n)\lambda t]$,
showing that the characteristic decay time $1/(1-n)\lambda$ of the dressed
Omori propagator $N(t)$
becomes much larger (much longer memory) that the decay time 
$1/\lambda$ of the bare
Omori propagator.

For $\mu>2$, we recover the
standard diffusion corresponding to $\theta >1$ and $\mu>2$ discussed
in section \ref{mgmls}.

\subsection{Long waiting times ($\theta < 1$) and long jump sizes
(L\'evy flight regime for $\mu \leq 2$)}

    Putting the leading terms of the expansions of ${\hat \Phi}(\vec k)$
and of ${\hat \Psi}(\beta)$ in (\ref{ngnslqw}) gives
\be
{\hat N}(\beta,\vec k) = {\hat S}_M(\beta,\vec k)~{1 \over
(\beta c')^{\theta} + (\sigma k)^{\mu}}~.
\label{ngnslqwaa}
\ee
The corresponding ${\hat W}(\beta,\vec k)$ is given by
\be
{\hat W}(\beta,\vec k) = {\hat S}_M(\beta,\vec k)~{(\beta)^{\theta
-1}  c'^{\theta} \over
(\beta c')^{\theta} + (\sigma k)^{\mu}}~.
\label{ngnslqaawaa}
\ee

Equation (\ref{ngnslqaawaa}) has been studied extensively in the context
of the CTRW model as a long wavelength $|\vec k| \to 0$ and long time
$\beta \to 0$
approximation to investigate the long time behavior of the CTRW.
Kotulski \cite{Kot1} has developed a rigorous approach, based on limit
theorems, to classify
the asymptotic behaviors of different type of CTRWs and justifies the
approximation
(\ref{ngnslqaawaa}) for the long time behavior. Barkai \cite{barkai2}
has studied
the quality of the long wavelength $|\vec k| \to 0$ and long time $\beta \to 0$
approximation (\ref{ngnslqaawaa}) by solving the exact CTRW problem for
the case when the waiting time distribution $\Psi(t)$ is a one-sided
stable L\'evy law of index $\theta$ with the same tail as (\ref{psidef})
and the distribution $\Phi(\vec r)$ of jumps is a symmetric
stable L\'evy of index $\mu$ with the same tail as (\ref{phidef}).
Their Laplace and Fourier transforms,
that appear in the denominator of (\ref{ngnslaaqw}), are
respectively ${\hat \Psi}(\beta) = \exp [-\beta^{\theta}]$ and
${\hat \Phi}(\vec k) = \exp [ -|\vec k|^{\mu}/2 ]$. Note that the
long wavelength $|\vec k| \to 0$ and long time $\beta \to 0$ approximation
gives $1-\exp [-(c' \beta)^{\theta}] ~\exp [ -|\sigma \vec k|^{\mu}]=
(c' \beta)^{\theta}+|\sigma \vec k|^{\mu}$, which recovers (\ref{ngnslqwaa}).
By comparing the exact solution of (\ref{ngnslqw}) for $\Psi(t)$ and
$\Phi(\vec r)$ of the above L\'evy form with that of
the long wavelength $|\vec k| \to 0$ and long time $\beta \to 0$
approximation (\ref{ngnslqaawaa}), Barkai \cite{barkai2} finds that certain
solutions of (\ref{ngnslqaawaa}) diverge on the origin, a behavior not found
for the corresponding solutions of (\ref{ngnslqw}). In addition, certain
solutions of the full equation (\ref{ngnslqw}) converge only very
slowly for $\mu <1$
to the solutions of the long-time approximation (\ref{ngnslqaawaa}). These
results validate our use of the asymptotic long time behavior with respect
to the scaling laws but provide a note of caution if one needs more precise
non-asymptotic information. In this case, such information can be obtained
by a suitable analysis of the full equation (\ref{ngnslqw}).

\begin{figure}
\includegraphics[width=16cm]{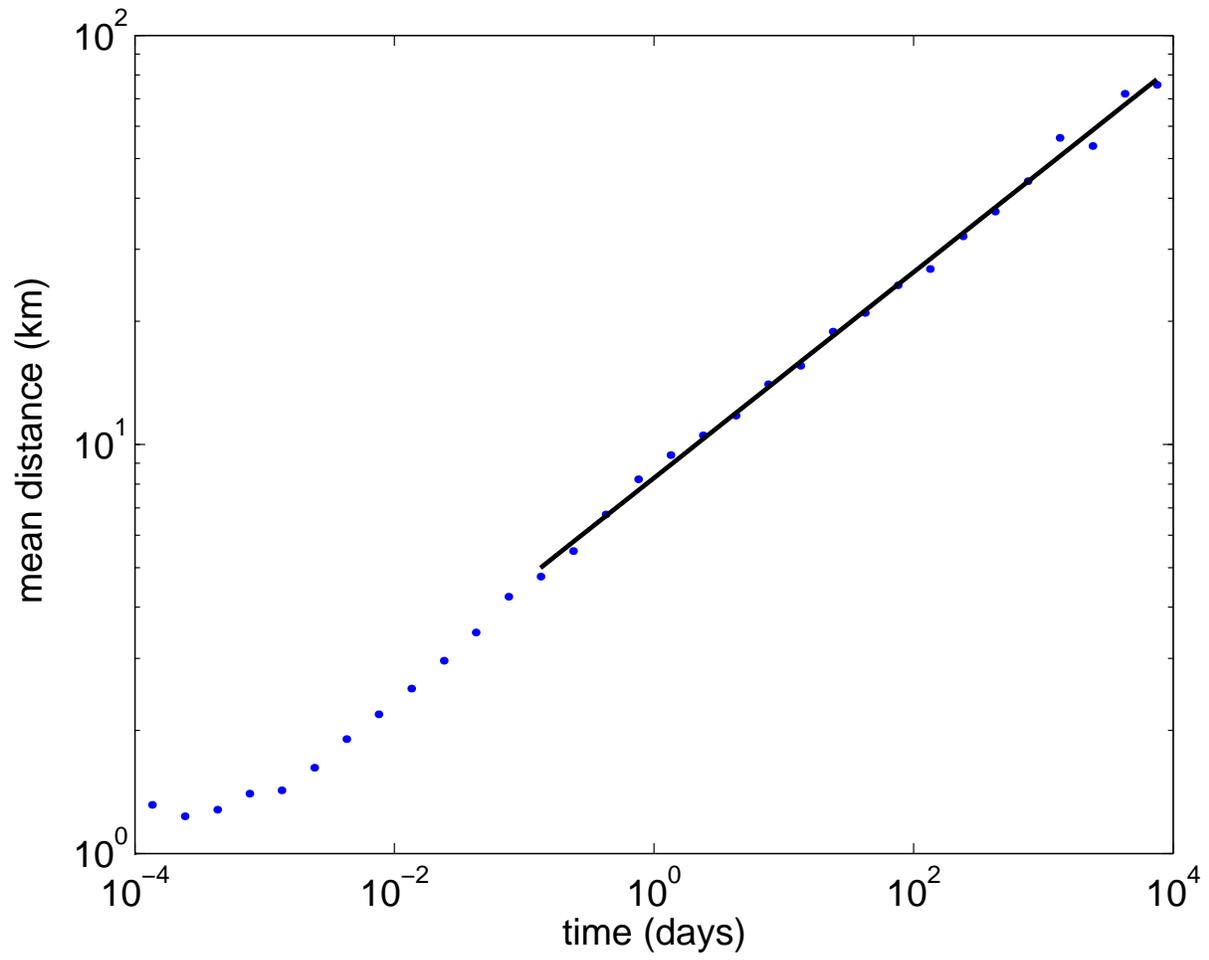}
\caption{\label{poimlksq} Average distance between the first mainshock and its
aftershocks as a function of the time from the mainshock, for a numerical
simulation of the ETAS model in the critical regime $n=1$, with
$\theta=0.2$, $\mu=0.9$, $c'=1$ day and $d=1$ km.
The solid line is a fit of the data which gives an exponent $H=0.25$ in good
  agreement with the predicted value $H=0.22$.}
\end{figure}

Using power counting,
expression (\ref{ngnslqaawaa}) predicts a diffusion process
(\ref{hgnvaa}) with exponent
\be
H = {\theta \over \mu}~.
\label{mgjgkr}
\ee
This prediction is checked by numerical simulation of the ETAS model
in the critical regime $n=1$, with $\theta=0.2$, $\mu=0.9$, shown in figure
\ref{poimlksq}. The average distance between the first mainshock and its
aftershocks as a function of the time from the mainshock indeed increases
according to (\ref{hgnvaa}) with an exponent $H$ in very
good agreement with the prediction $H=\theta/ \mu=0.22$. As
the form of the denominator in (\ref{ngnslqaawaa})  is
independent of the space dimension, the prediction
(\ref{mgjgkr}) is valid in any space dimension.

The natural variable for the expansions given below allowing to compute
$N(t, \vec r)$ is
\be
z={ D ~t^{\theta/\mu} \over |\vec r|}~,
\label{mgjgrk}
\ee
where $D=\sigma / c'^{\theta/\mu}$ and $c'=c 
\left(\Gamma(1-\theta)\right)^{1/\theta}$.

\subsubsection{$z$-expansion of the solution}

$W(t, \vec r)$ can be obtained as the following sum (equation (5.10)
of \cite{Saichev})
\be
W(t, \vec r) = {1 \over \pi  |\vec r|}~
\sum_{m=0}^{+\infty} (-1)^m~ z^{m \mu}
~{\Gamma(m \mu +1) \over \Gamma(m \theta +1)} ~\cos \left[ {\pi \over
2} (m \mu +1)\right]~.
\label{njhgrjb}
\ee

Applying (\ref{nbjn}) to (\ref{njhgrjb}) term by term in the sum, we get
\be
N(t, \vec r) = {c'^{-\theta} \over D \pi ~
t^{1-\theta+\theta/\mu}}~
\sum_{m=0}^{+\infty} (-1)^m~ z^{1+m \mu}
~{\Gamma(m \mu +1) \over \Gamma((m+1)\theta)} ~
\cos \left[ {\pi \over 2} (m \mu +1)\right]~,
\label{njhaagrssjb}
\ee
The asymptotics
\be
{\Gamma(m \mu+\mu +1)~\Gamma(m \theta +1)
\over \Gamma(m \theta + \theta +1)~\Gamma(m \mu +1)} \sim
{\Gamma(m \mu +\mu+1)~\Gamma((m+1)\theta) \over \Gamma((m+2)\theta)~
\Gamma(m \mu +1)}
\sim m^{\mu - \theta}
\label{mgjgl}
\ee
show that the series (\ref{njhgrjb})
and (\ref{njhaagrssjb}) exist only for $\mu < \theta$. It can be shown that
these series exist for all $z$ in this case.
This series converges very slowly for large $z$ but
the Pad\'e summation method \cite{BeOr} can be used to improve the convergence
of  (\ref{njhaagrssjb}) in the case $\mu < \theta$, and can also be used
  to estimate     (\ref{njhaagrssjb}) in the case $\mu > \theta$ for which the
  series diverges.

The space integral $\int dr ~N(t, r)$ over the whole one-dimensional
volume $V$,
with $N(t, r)$ given by (\ref{njhaagrssjb}),
recovers the global Omori law
\be
\int_V dr ~N(t, r) \sim  {1 \over t^{1 - \theta}}~.
\label{nhnkjbkf}
\ee
Note the non-trivial phenomenon in which the superposition of all
aftershock activities transforms
the local Omori law or ``bare propagator'' (\ref{psidef})
$\Psi(t) \sim  {1 \over t^{1+\theta}}$ into the global
Omori law or ``dressed propagator'' ${1 \over t^{1 - \theta}}$.
This effects was predicted in \cite{SS,HS1} in the version of the ETAS model
without space dependence.
These results are consistent with the claim of section
\ref{consosk} according to which all results
reported previously for the version
of the ETAS model without space dependence hold also for the version of the
space-dependent ETAS model studied here, when averaging over the whole space.

The asymptotic behavior for  $|\vec r| \gg D~ t^{\theta \over \mu}$
(i.e., $z \ll 1$)
and $\mu < \theta$  is obtained by keeping only the first non-zero
term ($m=1$) in (\ref{njhaagrssjb})
which is convergent for all $z$ in the case   $\mu < \theta$
\be
N(t, \vec r) = { \sin \left({\pi \mu \over 2}\right) \over \sigma
c'~\pi}~ {\Gamma(1+\mu)
\over \Gamma(2\theta)}~\left({c' \over t}\right)^{1-2 \theta}~
\left({\sigma \over |\vec r|}\right)^{1+\mu}~,~~~~{\rm for}~~
|\vec r| \gg D~t^{\theta \over \mu}~.
\label{gjjsls}
\ee
At fixed large $|\vec r|$ and for $t< |{\vec r}/D|^{\mu \over \theta}$,
this predicts a local Omori law with exponent $p=1-2\theta$.

\subsubsection{$1/z$-expansion of the solution}

We use the theory of Fox functions \cite{Mathai} to obtain
$N(t,\vec r)$ as an infinite series in $1/z$. For this, we first
rewrite expression (\ref{njhaagrssjb}) as a Fox
function \cite{Mathai}
\be
N(t,\vec r) = {c'^{-\theta} \over D ~\mu~ \pi~
    ~ t^{1-\theta+\theta/\mu}}~
R\left( H_{2,2}^{1,2}
\left[ z~e^{i \pi /2} \left |
{\begin{array}{l} (1/\mu, 1/\mu),(1,1) \\ (1/\mu, 1/\mu),
(\theta/\mu-\theta+1, \theta/\mu) \end{array}}
\right .
\right ]  \right)~,
\label{fox4}
\ee
where  $R(z)$ indicates the real part of $z$.

The $1/z$ expansion of  $N(t, \vec r)$ can be obtained using
the dual expansion of the Fox function  (\ref{fox4}) (expression
(3.7.2) of \cite{Mathai})
$$
N(t,\vec r) = {c^{-\theta} \over D ~ \pi~ \mu~
           ~ t^{1-\theta+\theta/\mu}}~
          \sum_{m=0}^{+\infty} (-1)^m ~\biggl[\mu~
         z^{1-\mu-m\mu}~{\Gamma(1-(m+1)\mu) ~\sin((m+1)\mu\pi/2)
          \over \Gamma(-m \theta )}
$$
\be
\left.  + {z^{-m} \over m!}~ {\pi   ~\cos(m \pi/2)
          \over  \sin((m+1)\pi/\mu) ~ \Gamma(\theta -(m+1)\theta/\mu )} \biggl]
           \right.
\label{fox5}  ~.
\ee
This expansion exists only for $\mu>\theta$ (conditions of page 71 below
eq.~(3.7.2) of \cite{Mathai}). This is easily checked by the behavior
of an asymptotics similar to (\ref{mgjgl}).  Note that the series (\ref{fox5})
is not defined in the
special case $\mu=1$ due to the presence of the ill-defined ratio
$\Gamma(0)/\Gamma(0)$ and a different approach is required, such as the
integral representation of $W(t,\vec r)$ developed in \cite{Saichev}.
The global Omori law obtained by integrating over the whole space
(\ref{fox5})
  is again $N(t) \sim 1/t^{1-\theta}$ as expected from the analysis of
the ETAS model
  without space dependence \cite{HS1}.

Keeping only the largest term of  (\ref{fox5}) for large $z$, we
obtain the asymptotic
  behavior for small distances  $ r < D~t^{\theta /\mu}$
  \ba
  N(t,r) &\simeq&
  {\Gamma(1-2\mu) ~
   ~\sin(\pi \mu) ~\sin(\pi \theta) ~ \over  c' \sigma~ \pi^2}
  {\Gamma(1+\theta) \over  (r/\sigma)^{1-2\mu}}~{1 \over (t/c')^{1+\theta}}
  ~~~\rm{for}~~ \mu<0.5  \nonumber \\
  N(t,r) &\simeq&   { c'^{-\theta}  \over
         c' \sigma~\mu~\Gamma(\theta-\theta/\mu) ~\sin(\pi/ \mu)}~
         {1 \over (t/c')^{1-\theta+\theta/\mu}}
     ~~~~~~~~~~~~~~~~\rm{for}~~ 0.5<\mu<2 ~.
  \label{lqps1}
  \ea
  Note that for $ r < D~t^{\theta /\mu}$ and $0.5<\mu<2$, the leading behavior
  of $N(t,r)$ is independent of $r$.

\begin{figure}
\includegraphics[width=12cm]{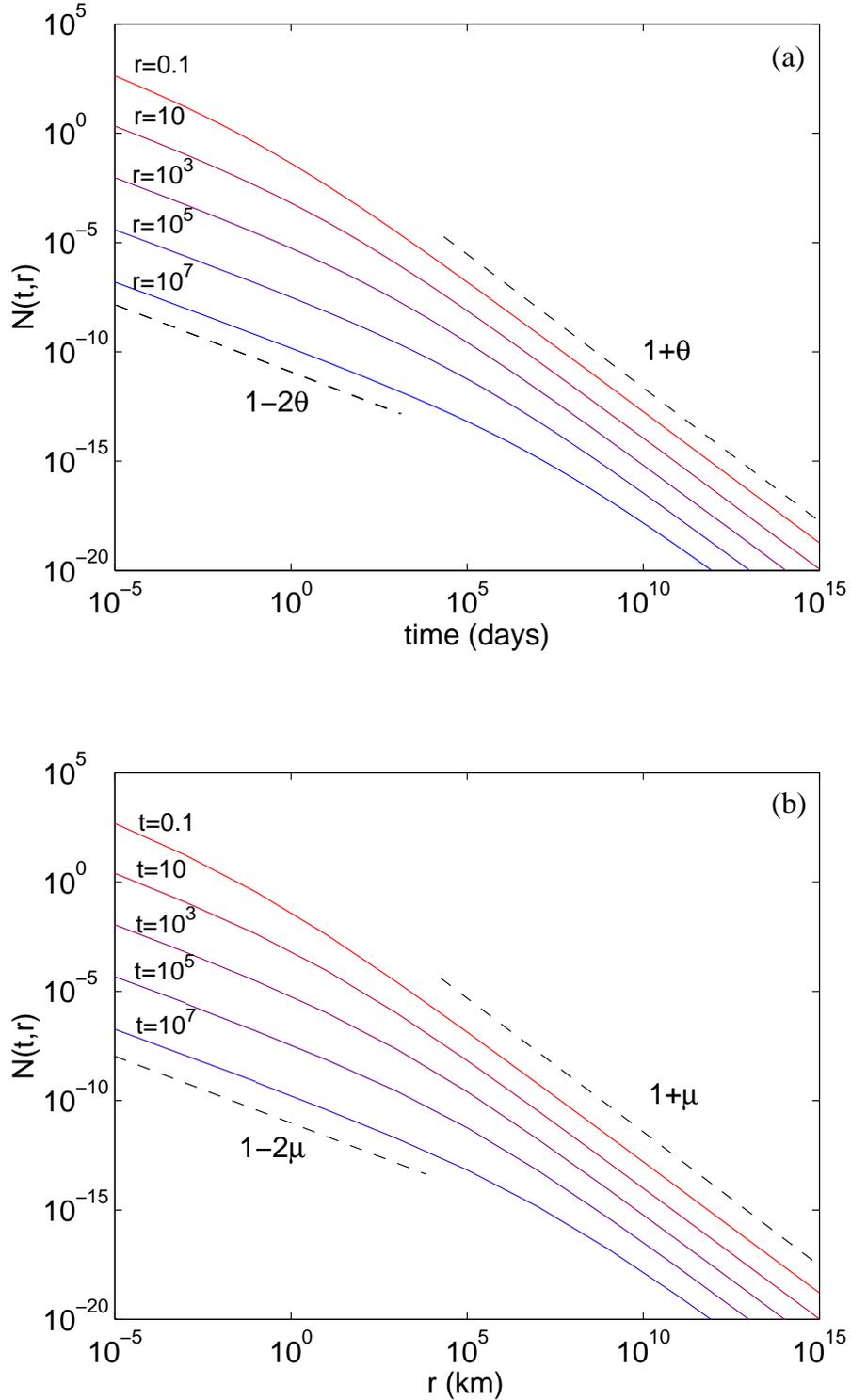}
\caption{\label{piuonk} Rate of seismicity $N(t,r)$ for $\theta=0.2$,
$\mu=0.2$,
  $c'=1$ day and $\sigma=1$ km,
evaluated from expressions (\ref{njhaagrssjb}) and (\ref{lqps1}),
  plotted as a function of the time (a) for different values of the distance
  $r$ between the mainshock and its aftershocks, and (b) as a function of $r$
  for different values of the time between the mainshock and its aftershocks.
We stress again that the time scales shown here do not necessarily correspond
to real observable time scales but are presented to demonstrate clearly the
existence of the two regimes. The dashed lines give the predicted
asymptotic dependence in each regime.
}
\end{figure}

\begin{figure}
\includegraphics[width=12cm]{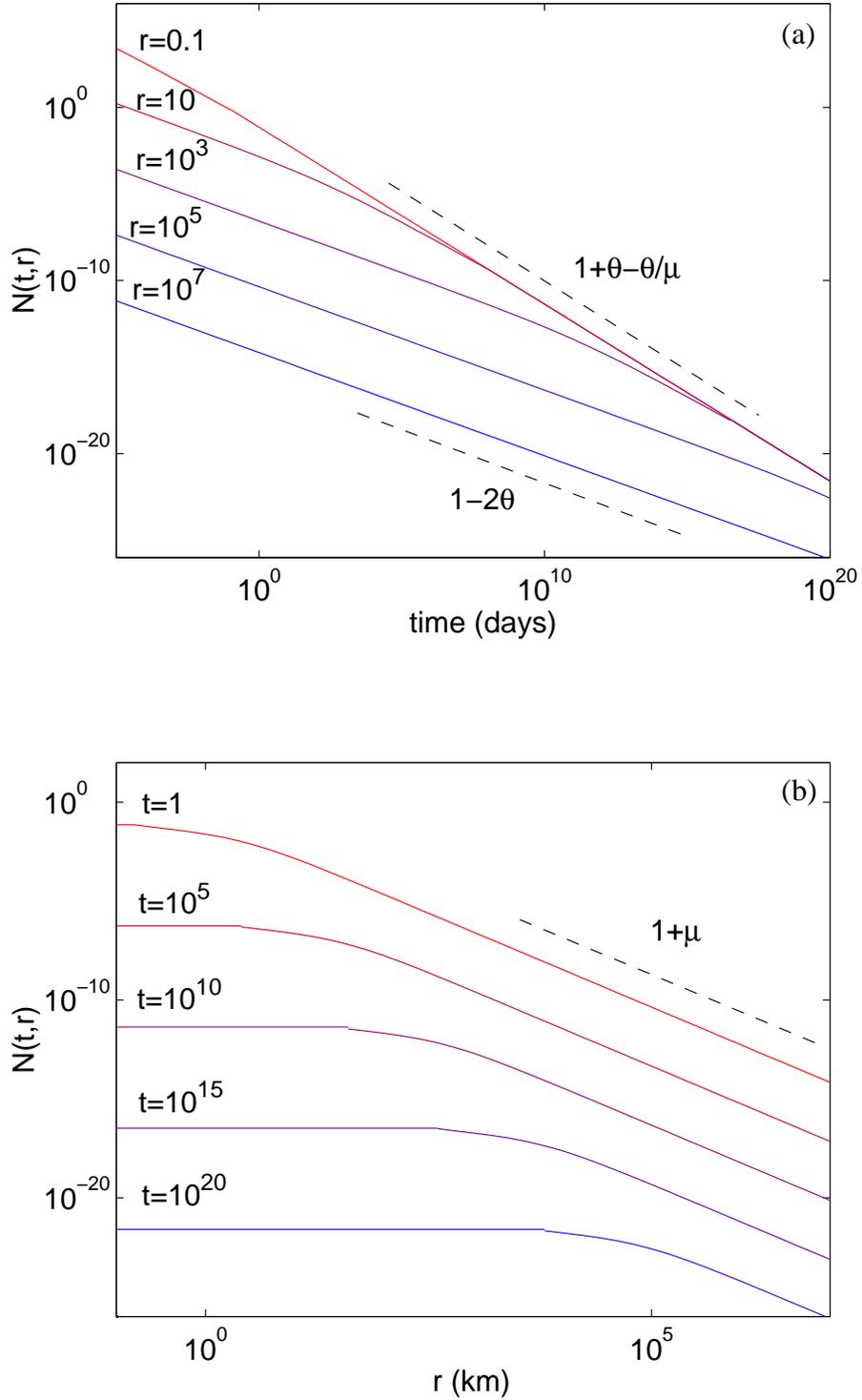}
\caption{\label{piuon} Rate of seismicity $N(t,r)$ for $\theta=0.2$, $\mu=0.9$,
  $c'=1$ day and $\sigma=1$ km, evaluated from expressions (\ref{njhaagrssjb}) and (\ref{lqps1}),
  plotted as a function of the time (a) for diffe
rent values of the distance $r$
   between the mainshock and its aftershocks, and (b) as a function of
$r$ for different
    values of the time between the mainshock and its aftershocks.
   The dashed lines give the predicted
asymptotic dependence in each regime.
    }
\end{figure}

Equation (\ref{lqps1}) thus predicts an apparent exponent
\ba
   p &=& 1+\theta        ~~~~~~~~~~~~~~~~\rm{for}~~ \mu<0.5  \nonumber \\
   p &=& 1-\theta +  \theta/\mu ~~~~~\rm{for}~~ 0.5<\mu<2
\label{leqps5}
\ea
  for small distances $r < D~t^{\theta /\mu}$.
This prediction is valid only in the case $\mu>\theta$ for which the series
(\ref{fox5}) is convergent.
However, the same asymptotic results are also obtained by different methods
in the case  $\mu<\theta$, for instance
expression (\ref{leqps5}) is recovered for all $\mu<2$ using the
integral representation  of \cite{Saichev} [A. Saichev, private communication].
The numerical evaluation of (\ref{njhaagrssjb}), which converges for
  $\mu < \theta$, also recovers the asymptotic results (\ref{lqps1}).
The two regimes $\mu<0.5$ and $0.5<\mu<2$ are illustrated in Figures
\ref{piuonk}
  and \ref{piuon} respectively. The seismicity rate $N(t,\vec r)$ is evaluated
  from expression (\ref{njhaagrssjb}) for small $z$ and from 
expression (\ref{fox5})
  for large $z$.

\begin{figure}
\includegraphics[width=10cm]{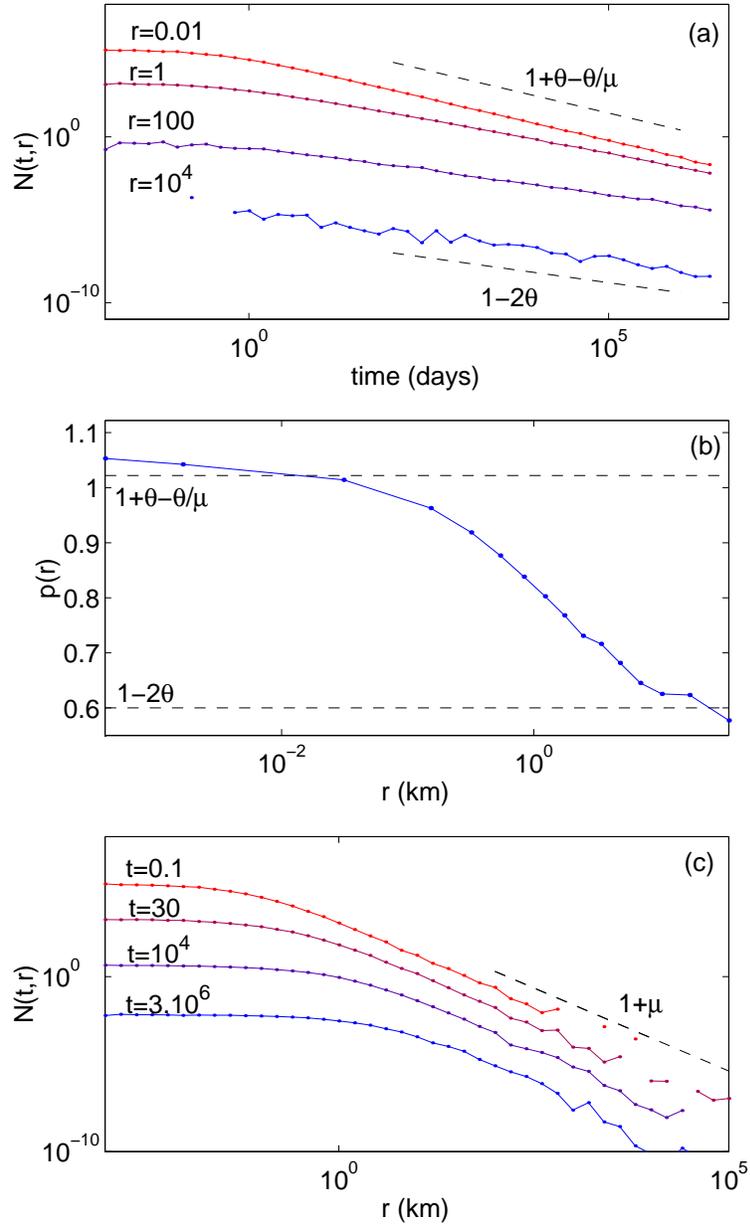}
\caption{\label{poiq}
  Rate of seismicity $N(t,r)$ obtained from numerical simulations of
the ETAS model generated
  with the same parameters as in Figure \ref{piuon} ($\theta=0.2$, $\mu=0.9$,
  $c'=1$ day and $d=1$ km).
$N(r,t)$ is computed by averaging over 500 numerical
realizations of the ETAS model. (a)  aftershock rate as a function of the
  time from the mainshock for several distances
$|\vec r|$ ranging from $0.01$ to $10^4$ km. (b)
Apparent Omori exponent measured for times $t>10$
as a function of the distance from the mainshock.
The aftershock decay rate (with time) is larger close to the
mainshock epicenter
than at large distances from the mainshock. The asymptotic values for small and
large distances are in agreement with the predictions (\ref{leqps5}) for
  $r \ll D t^{\theta/\mu}$ and (\ref{gjjsls}) for $r \gg D
t^{\theta/\mu}$, which
  are shown as the horizontal dashed lines. (c) Rate of seismicity $N(t,r)$
  as a function of the distance between aftershocks and mainshock for various
  times. The theoretical prediction for large distances is shown as
the dashed line
  with slope $-(1+\mu)$.
  }
\end{figure}

We also performed numerical
simulations of the ETAS and CTRW models and the results are in good 
agreement with
expression (\ref{njhaagrssjb}) and (\ref{fox5}) for $N(\vec r,t)$ for $t \gg c$
  and $r \gg d$. For very small times $t\ll c$, or for very small distances
  $r \ll d$, expressions (\ref{njhaagrssjb}) and (\ref{fox5}) are not 
valid because
  they are based on a long wavelength $|\vec k| \to 0$ and long time 
$\beta \to 0$
  approximation.
Numerical simulations of the ETAS model in the case $\theta=0.2$ and $\mu=0.9$
are presented in Figure \ref{poiq}, and are in good agreement with 
the analytical
solutions (\ref{njhaagrssjb}) and (\ref{fox5}) shown in Figure \ref{piuon}
for the same parameters, except from the truncation of $N(t,r)$ for times
$t\ll c$ and distances $r \ll d$ that are not reproduced by the 
analytical solution.

\subsection{A simple non-separable joint distribution
of waiting times and jump sizes: coupled spatial diffusion and long
waiting time distribution}

Consider the choice for $\phi_{m_i}(t-t_i, \vec r-\vec r_i)$
replacing (\ref{first})
by
\be
\phi_{m_i}(t-t_i, \vec r-\vec r_i) = \rho(m_i)~\Psi(t-t_i)~\Phi(|\vec
r-\vec r_i|/\sqrt{D t})~,
\label{firsaat}
\ee
where $\rho(m_i)$ and $\Psi(t)$ are again given by (\ref{formrho})
and (\ref{psidef})
while (\ref{phidef}) is changed into
\be
\Phi(|\vec r-\vec r_i|/\sqrt{D t}) = {1 \over \sqrt{2Dt}} ~
\exp \left(-|\vec r-\vec r_i|^2/D t\right)~.
\label{mgmglw}
\ee
The spatial diffusion of seismic activity is now coupled to the
waiting time distribution. Expression (\ref{mgmglw}) captures the
effect that, in order for aftershocks to spread over large distances
by the underlying physical process, they need time. In fact,
returning to the discussion in the introduction on the various
proposed mechanisms for aftershocks, expression (\ref{mgmglw})
embodies a microscopic diffusion process.

In this case, (\ref{ngnslqw}) must be replaced by
\be
{\hat N}(\beta,\vec k) = {{\hat S}_M(\beta,\vec k) \over
1 - n {\hat \phi}(\beta, \vec k)}~,
\label{ngssnslqw}
\ee
where ${\hat \phi}(\beta, \vec k)$ is the Laplace-Fourier transform
of the product
$\Psi(t)~\Phi(|\vec r|/\sqrt{D t})$. For large times and long distances for
which the first terms in the expansion in $\beta$ and $k$ are sufficient, and
for $n=1$, we obtain
\be
{\hat \phi}(\beta, \vec k) \propto {{\hat S}_M(\beta,\vec k) \over
(\beta + D k^2)^{\theta}}~.
\label{gjjgws}
\ee
The inverse Laplace-Fourier transform of (\ref{ngssnslqw}) is
\be
N(t, \vec r) \sim {1 \over t^{1-\theta}}~{1 \over \sqrt{2 \pi Dt}}~
    \exp\left(-|\vec r|^2/D t\right)~.
    \label{mjgwlw}
\ee
As expected, expression (\ref{mjgwlw}) recovers the dressed Omori propagator
in the case of absence of space dependence \cite{HS1}. At finite $r$
and long times,
the dressed Omori law also decay as $1 /t^{1-\theta}$. The diffusion of
aftershocks is normal with the standard diffusion exponent $H=1/2$.

\section{New questions on aftershocks derived from the CTRW analogy}

We list a series of comments and questions suggested from the analogy
between the ETAS model and the CTRW model. In particular, we discuss the
possibility of defining new observables for earthquake aftershocks, that could
be worthwhile to investigate in future empirical studies of
earthquake aftershocks.

\subsection{Recurrence of aftershock activity in the proximity of the
main shock}

A quantity often investigated in studies of random walks is the
probability $W(t, \vec 0)$
to find the random walker at its starting point (the origin) at time $t$.
In the earthquake framework, this is the seismic aftershock rate
close to the main shock.

\subsection{First-passage times}

The first passage time of a random walk is the first arrival time
of the random walk at a given point $\vec r$. In the earthquake context,
this translates into the study of the waiting time for a given region
to have its own first aftershock after the main shock occurs.
The distribution of such first passage waiting times gives the distribution
of times with no nearby seismic activity. See for instance \cite{barkai1}
in the case of a power law distribution of waiting times and Gaussian
distribution of jump sizes. Margolin and Berkowitz \cite{Margolin} give
the distribution of first-passage times in the case where the jump distribution
is narrow and the waiting distribution is long-tailed $\sim 1/t^{1+\theta}$.
They analyze the three different regimes $\theta <1$, $1<\theta <2$ 
and $\theta \geq 2$.

\subsection{Occupation time of seismic activity}

Weiss and Calabrese \cite{Weiss2} have studied the total amount of time spent
by a lattice CTRW on a subset of points. In the seismic language,
this amounts to study the probability distribution of the durations
of aftershock sequences that are localized in a specific subset of the space.
In other words, how probable are aftershock sequences that are found only
within a given spatial subset over a certain duration?

\subsection{Transience and recurrence of seismic activity}

Another question that has been studied in some details in the CTRW framework
is whether random walks are transient or recurrent. A transient random walk
visits any point $\vec r$ at most a finite number of times before escaping to
infinity. For earthquakes, the transient regime corresponds to the activation
of at most a finite number of aftershocks in any given point $\vec r$.
In contrast, a recurrent random walk may return a growing number of times
to all or a subset of points at time increases. In the aftershock language,
this means that these points will have a
never-ending (decaying) aftershock activity. We stress here the difference
between the global Omori law giving a never-ending power law decay of
the aftershock activity (in the sub-critical regime $n<1$) and its
spatial dependence which must exhibit important variations. In particular,
in the recurrent regime, an Omori law can be documented by counting
aftershocks in those limited regions of space which are activated
again and again.

\subsection{Probability for the cumulative  number of aftershocks}

Let us define a basic quantity in the CTRW formalism, namely the probability
$\chi_m(t)$ to make exactly $m$ steps up to time $t$. In the
earthquake context,
$\chi_m(t)$ is the probability to have exactly $m$ aftershocks after
the main shock.
In the case in which the spatial transition probability $\Phi(\vec r)$ between
different positions is independent of the waiting times (corresponding to
factorizing $\phi_{m_i}(t-t_i, \vec r-\vec r_i)$ as in (\ref{first})), the
probability density $W(t, \vec r)$ to find the walker at position $\vec r$ at
time $t$ can be written
\be
W(t, \vec r) = \sum_{m=0}^{+\infty} W_m(\vec r)~\chi_m(t)~,
\label{gkmkvms}
\ee
where $W_m(\vec r)$ is the probability to reach $\vec r$ from $\vec
0$ in $m$ steps.
In the earthquake context, $W_m(\vec r)$ is the probability that
there has been exactly
$m$ events in the time interval $[0,t]$ and that the last one
occurred at $\vec r$.
Equation (\ref{gkmkvms}) states that the CTRW is a random process
subordinated to
simple random walks described by $W_m(\vec r)$
under the operational time given by the $\chi_m(t)$ distribution
\cite{Sokolov1,Sokolov2}.

\subsection{Random walk models with birth and death and background
seismicity from localized sources}

Bender et al. \cite{Bender} have studied models of random walks in
which walkers are born
in proportion to the population
at one specific site (for instance the origin)
with probability $a-1$ (with $a>1$) and die at all
other sites with probability $1-n$ (with $n \leq 1$). In the
earthquake context, this consists in assuming that the aftershock activity is
fed by a localized region in space, which is itself activated by the aftershocks
returning to this region, furthering the overall activity. This may
be considered
to describe the seismic activity close to a plate boundary, in which the plate
boundary is the constant self-consistent source of a seismic activity
which may spread over
a significant region away from the boundary. The excursion of the
random walkers
quantify the spread of the seismic activity away from the main fault structure.
The rate of death of the walkers correspond exactly to the distance $1-n$ from
the critical value $n=1$. Bender et al. \cite{Bender} find a phase
diagram in the
$(a-1, 1-n)$ parameter space in which a boundary separates two possible
asymptotic regimes:
\begin{enumerate}
\item for small $a-1$ and large $1-n$, the seismic activity at the origin
and everywhere eventually dies off;
\item for large $a-1$ and small $1-n$, the average seismic activity
at the origin approaches
a positive constant at long times. In this regime, there is a
transition as $a-1$
is decreased or as $1-n$ is increased, between
a case where the global seismic activity outside the origin goes to
zero and a case where it
diverges at long times. On the boundary between these two regimes in the
$(a-1, 1-n)$ parameter space, the distribution of seismic activity
approaches a steady state
at long times. There is a critical point (for space dimensions
different from $2$)
at a certain value $(a_c-1, 1-n_c)$, for
which the long-time seismic activity away from the source is given by
$\sim (a - a_c)^{\nu}$
where $\nu$ is a critical exponent equal to $2$ in three dimensions.
\end{enumerate}

Note that the results of \cite{Bender} are obtained for random
walks on a lattice.
This can easily be converted into a CTRW by the fact that
a CTRW is nothing by a process subordinated to
discrete random walks under the operational time defined by the
process $\{t_i\}$
of the time of just arrival to a given site, as given by (\ref{gkmkvms}).

\section{Discussion}

Using the analogy between the ETAS model and the CTRW model
established here, we have derived the relation between the average
distance between aftershocks and the mainshock as a function of the
time from the mainshock, and the joint probability distribution
of the times and locations of aftershocks.

We have assumed that each earthquake triggers aftershocks at a distance $r$
and time $t$ according to the bare propagator $\phi(r,t)$, which can
be factorized as $\Psi(t)\Phi(r)$. This means that the distribution
$\Phi(r)$ of the distances between an event and its direct aftershocks is
decoupled from the distribution  $\Psi(t)$ of waiting time. Hence, the direct
aftershocks triggered by a single mainshock do not diffuse in space with time.
Notwithstanding this decoupling in space and time of the bare propagator
$\phi(r,t)$, we have shown that the global law  or dressed propagator
$N(t,\vec r)$ defined as the global rate of events at time $t$ and at position $\vec r$,
cannot be factorized into two distributions of waiting times and space jumps.
This joint distribution of waiting times and positions of the whole sequence of
aftershocks cascading from a mainshock is different from the product 
of the bare time and space propagators.

\begin{figure}
\includegraphics[width=10cm]{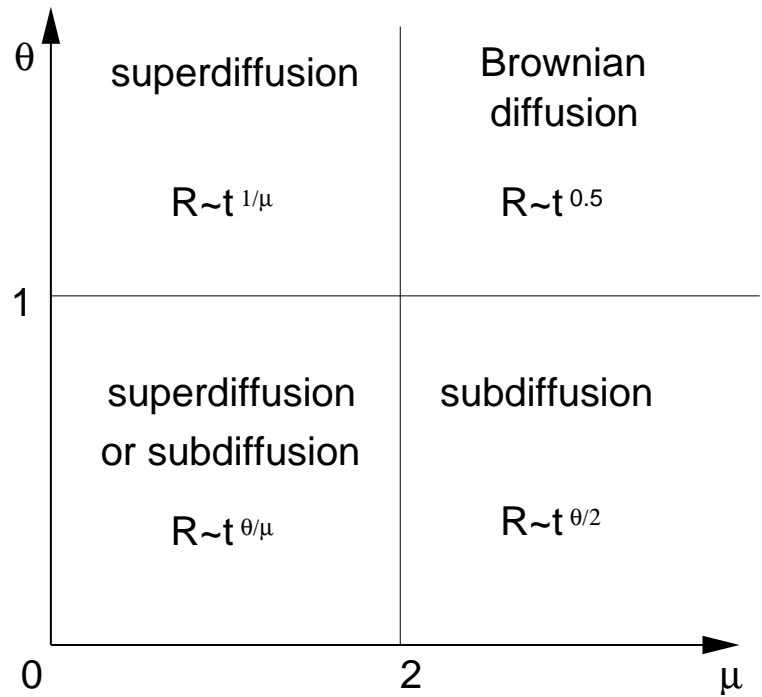}
\caption{\label{H} Classification of the different regime of the diffusion
of aftershocks in space as a function of time from the main shock.
The bare Omori law for aftershocks decay with time as $1/t^{1+\theta}$.
The jump size distribution between the earthquake
``mother'' and its ``daughters'' is proportional to
$1/r^{1+\mu}$. $R(t)$ is the average distance between all
aftershocks triggered up to time $t$ after the mainshock.}
\end{figure}

The mean distance between the mainshock and its aftershocks, including
secondary aftershocks, increases with the time from the mainshock,
due to the cascade process of aftershocks triggering aftershocks triggering
aftershocks, and so on.
In the critical case $n=1$, this diffusion takes the form
of a power-law relation $R \sim t^H$ of the average distance $R$ between
aftershocks and the mainshock, as a function of the time $t$ from the
mainshock.
If the local Omori law is characterized by an exponent $0<\theta<1$, and if
the space jumps follow a power law $\Phi(r) \sim 1/(r +d )^{1+\mu}$,
the diffusion exponent is given by $H=\theta/\mu$ in the case $\mu<2$ and
$H=\theta/2$ in the case $\mu>2$. Depending on the $\theta$ and $\mu$ values,
we can thus observe either sub-diffusion ($H<1/2$) or super-diffusion
($H>1/2$), as summarized in Figure \ref{H}.
In the sub-critical ($n<1$) and super-critical ($n>1$) regimes,
this relation is still valid up to the characteristic time $t^*$
given by (\ref{tstar})
and for distances smaller than $r^* \propto D t^{*H}$ given by (\ref{gnjgrkd}).
For $t>t^*$ and $r>r^*$ in the sub-critical regime,
the global distributions of times and distances
between the mainshock and its aftershocks are decoupled and there is therefore
no diffusion. In the super-critical regime, the aftershock rate increases
exponentially for $t>t^*$ and the aftershocks diffuses more rapidly
than before $t^*$.

In the critical regime, the cascade of secondary aftershocks introduces a
variation of the apparent Omori exponent as a function of the
distance from the mainshock. The asymptotic values of the Omori exponent
in the different regimes are summarized in Table \ref{table2}.
In the regime $\mu<2$, we observe a transition from an Omori law decay
with an exponent $p=1-2\theta$ at early times $t^H \ll r/D$ to
a larger exponent at large times. This provides
another mechanism to explain the observed variability of the Omori exponent.
In the regime $\mu>2$, a power-law decay of the seismicity with time is
observed only at large times $t^H \gg r/D$. At early times, or at large
distances $r \gg D t^H$, the seismicity rate is very small, because the
seismicity as not yet diffused up to the distance $r$.

We should emphasize that our theoretical analysis of aftershock
diffusion predicts the behavior of the ensemble average of aftershock sequences.
Individual
sequences may depart from this ensemble average, especially for 
sequences with few earthquakes and limited durations. For long sequences
(20,000 events say), we have verified that the exponent $H$
measured on individual sequences does not deviate from the ensemble
average value by more than about 20\%. As already discussed, the
impact of fluctuations become however more effective as the parameter $\alpha$
increases above $b/2$.

The diffusion of the seismicity also renormalizes the spatial distribution
of the seismicity, which is very different from the local
distribution $\Phi(r)$
of distances between a triggering event and its direct aftershocks.
In the regime $\mu>2$, the global seismicity rate $N(t,\vec r)$ decays
exponentially with the distance from the mainshock, whereas the local
distribution of distances $\Phi(r)$ is a power-law distribution.
In the regime $\mu<2$, the local law $\Phi(r) \sim r^{-1-\mu}$ is
recovered at large distances, but a slower decay for $0.5<\mu<2$ or a constant
rate for $\mu<0.5$ is observed at small distances $r \ll D t^H$.
These predictions on the decrease of the Omori exponent with $r$
have not yet been observed in earthquake catalogs, but
an expansion of the aftershock zone has been reported in many studies
\cite{Mogi,Imoto,Chatelain,Tajima1,Tajima2,Wesson,Ouchi,Noir,Jacques}.
However, very few studies
have quantified the diffusion law. Noir et al. \cite{Noir}
show that the earthquake Dobi sequence (central Afar, August 1989)
composed of 22 $M>4.6$ earthquakes presented a migration that was
in agreement with a diffusion process due to fluid transfer in the crust,
characterized by a normal diffusion process with exponent $H=0.5$.
Tajima and Kanamori \cite{Tajima1,Tajima2} studied several aftershock
sequences in subduction zone and observed a much slower logarithmic
diffusion, which is compatible with a low exponent $H$ close to $0.1$.
In some cases, the aftershock sequence displays no expansion with time.
For instance, Shaw \cite{Shaw} studied several aftershock sequences in
California and concluded that the distribution of distances between the
mainshock and its aftershocks is independent of time.
This can be explained by the fact that the 0mori exponent  measured
in \cite{Shaw} is very close to $1$, thus $\theta$ is very small and our
prediction is that the exponent $H$ should be very small.

In fact, the ETAS model predicts that diffusion should be observed only
for aftershock sequences with a measured Omori exponent $p$ significantly
smaller than $1$, which can only occur according to our model when the bare
Omori propagator with exponent $1+\theta$ is renormalized into the
dressed propagator with global exponent $1-\theta$. We have shown that
this renormalization of the exponent only occurs at times less than $t^*$,
while for longer times in the sub-critical regime $n<1$ the dressed Omori
propagator recovers the value of the bare exponent $1+\theta > 1$
(see figure \ref{n}).
Therefore, identifying an empirical observation of $p<1$ with our
prediction $p=1-\theta$ indicates that the aftershock sequence falls in the
 ``good'' time window $t<t^*$ in which the renormalization operates.
We have also shown that the dressed propagator gives a diffusion only for
$t<t^*$. We can thus conclude that, according to the ETAS model, the
observation of an empirical Omori
exponent larger than $1$ is indicative of the large time $t>t^*$ behavior
in the sub-critical regime $n<1$, for which there is no diffusion.
This provides a possible explanation for why many sequences studied by
\cite{Tajima1,Tajima2,Shaw} do not show a diffusion of the aftershock
epicenters.
Reciprocally, a prerequisite for observing diffusion in
a given aftershock sequence is that the empirical
$p$-value be less than $1$ in order to qualify the regime $t<t^*$.

An alternative model has been
discussed by Dieterich \cite{Diete} who showed that the spatial
variability of the stress
induced by a mainshock, coupled with a rate and state friction law, results
in an expansion of the aftershock zone with time. This expansion does not take
the form of a diffusion law as observed in the ETAS model,
the relation between the characteristic size
of the aftershock zone does not grow as a power law of the time from the
mainshock (equation (22) and Figure 6 of \cite{Diete}).

Marsan et al. \cite{Marsan1,Marsan2} and Marsan and Bean \cite{Marsan3}
studied several catalogs at different scales, from the scale of a
deep  mine to the world-wide seismicity, and observed that the average
distance between two earthquakes increases as a power-law of the time between
them, with an exponent often close to 0.2, indicative of a sub-diffusion process.
They interpreted their results as a mechanism of stress diffusion, that may
be due to fluid transfer with heterogeneous permeability leading to sub-diffusion.
Their analysis is quite different from those used in other studies, because
they consider all pairs of events, without distinction between
aftershocks and mainshocks.
This analysis can however lead to spurious diffusion, and in some
cases this method  does not detect diffusion in synthetic data set
with genuine diffusion.
We have tested their analysis on a synthetic catalog generated by superposing
a background seismicity with uniform spatial and temporal distribution, and
10 mainshocks with poissonian distribution in time and space, and with a
power-law distribution of energies. Each of these mainshocks generates only
{\it direct} aftershocks, without secondary cascades of aftershocks,
and the number of aftershocks increases exponentially with the
magnitude of the mainshock.
This way, we generate a synthetic catalog without any physical
process of diffusion, and which includes all the other well-established
characteristics of real seismicity: clustering in space and time superposed
to a seismicity background.
Applying the analysis of \cite{Marsan1,Marsan2,Marsan3} to this
synthetic data set leads to an apparent
diffusion process with a well-defined exponent $H=0.5$.
However, this apparent diffusion does not
   reflect a genuine diffusion but simply describes the crossover
from the characteristic size of an aftershock zone at early times to the
larger average distance between uncorrelated events at large times. In plain
words, the apparent power law $R \propto t^H$ is nothing but a cross-over
and is not real.
Furthermore, applying this analysis to a synthetic catalog generated
using the ETAS model, without seismicity background, and with a theoretical
diffusion exponent $H=0.2$, the method yields $H=0.01$ if we use all the
events of the catalog. If we select only events up to a maximum
distance $r_{max}$ to apply the same procedure as in
\cite{Marsan1,Marsan2,Marsan3},
we obtain larger values of $H$ which are more in agreement with the
theoretical exponent  $H=0.2$ but with large fluctuations that are function of
$r_{max}$.
Therefore, it is probable
that the diffusion reported in \cite{Marsan1,Marsan2,Marsan3} is not real
and results from a cross-over between two characteristic
scales of the spatial earthquake distribution. It may be attributed to
the analyzing methodology which mixes up uncorrelated events.
We are thus reluctant to compare the results of Marsan et al.
\cite{Marsan1,Marsan2,Marsan3}
with the predictions obtained with the ETAS model.

One can similarly question the results on anomalous diffusion of seismicity
obtained by Sotolongo-Costa et al. \cite{Sotolongo}, who
considered 7500 micro-earthquakes recorded by a local spanish network
from 1985 to 1995. They interpret the sequence of earthquakes
as a random walk process, in which the walker jumps from an
earthquake epicenter
to the next in sequential order. The time between two successive events is
seen as a waiting time between two jumps and the distance between these
events is taken to correspond to the jump size. Since the distributions of
time intervals and of distances between successive earthquakes are both
heavy-tailed (approximately power laws), their model is a CTRW.
We cannot stress enough that their CTRW model of seismicity has nothing to
do with our results on the mapping of the ETAS model onto a CTRW.
Their procedure is ad-hoc and their results depend obviously strongly on
the space domain of the analysis since distant earthquakes that are
completely unrelated can be almost simultaneous!
We also stress that our mapping of the ETAS model onto the CTRW model
does not correspond to identifying an earthquake sequence as a {\it single}
realization of a CTRW, as assumed arbitrarily by Sotolongo-Costa et al.
\cite{Sotolongo}.

Our predictions obtained here are thus difficult to test on
seismicity data, due to the small
  number of events available and the restricted time periods and
distance ranges,
  and because the seismicity background can strongly affect the results.
New methods should hence be developed to investigate if there is a
real physical
process of diffusion in seismic activity and to
compare the observations of real seismicity with the quantitative predictions
of the ETAS model.
Preliminary study of aftershock sequences in California leads to the conclusion
that most aftershock sequences are characterized by an Omori exponent $p>1$,
indicative of the sub-critical regime with $t>t^*$. As expected from our
predictions in this regime, we do not observe
an expansion of the aftershock zone.
However, a few sequences give a value $p<1$ and also exhibit an
increase of the average
distance between the mainshock and its aftershocks consistent with our
predictions. A detailed report of this analysis will be reported elsewhere.

\section{Conclusion}

We have studied analytically and numerically the ETAS
(epidemic-type aftershock) model, which is a simple
stochastic process modeling seismicity, based on the two best-established
empirical laws for earthquakes, the power law decay of seismicity
after an earthquake and a power law distribution of earthquake energies.
This model assumes that each earthquake can trigger aftershocks, with
a rate increasing with its magnitude.
In this model, the seismicity rate is the result of the whole cascade of
direct and secondary aftershocks.

We have first established an exact correspondence between the ETAS model
and the CTRW (continuous-time random walk) model. We have then used
this analogy to derive the joint probability of times and distances of the
seismicity following a large earthquake and we have characterized
the different regimes of diffusion.

We have shown that the diffusion of the seismicity should be observed only
for times $t<t^*$, where $t^*$ is a characteristic time depending on the model
parameters, corresponding to an observed Omori exponent smaller than one.
Most aftershock sequences have an observed Omori exponent larger than one,
corresponding to the subcritical regime of the ETAS model, for which there
is no diffusion.
The diffusion of the seismicity produces a decrease of the Omori
exponent as a function of the distance from the mainshock, the decay
of aftershocks being faster close to the mainshock than at large distances.
The spatial distribution of seismicity is also renormalized by the
cascade process,
so that the observed distribution of distances between the mainshock
and its aftershocks can be fundamentally different from the bare
propagator $\Phi(r)$ which gives the distribution of the distances
between triggered and triggering earthquakes. We have also
noted that the ETAS model generates apparent but realistical fractal
spatial patterns.

Assuming that the distances between triggering and triggered events
are independent of the time between them, this model generates a diffusion of
the whole sequence of aftershocks with the time from the mainshock,
which is induced by the cascade of aftershocks triggering aftershocks, and so on.
Our results thus provides a simple explanation of the diffusion of
aftershock sequences
reported by several studies, which was often interpreted as a mechanism
of anomalous stress diffusion. We see that no such ``anomalous stress
diffusion'' is
needed and our theory provides a parsimonious account of aftershock diffusion
resulting from the minimum physical ingredients of the ETAS model.
As Einstein once said, ``A theory is more impressive the greater
the simplicity of its premises, the more different the kinds of things
it relates and the more extended its range of applicability.''

\acknowledgments
We are very grateful to B. Berkowitz,  S. Gluzman, J.-R. Grasso, Y. Klafter, L. Margerin,
A. Saichev and  G. Zaslavsky for useful suggestions and discussions.

\newpage

\begin{table}[]
\begin{center}
\begin{tabular}{|c|c|c|}
     & ETAS & CTRW  \\ \hline
     $\Psi(t)$ & pdf for a ``daughter'' to be born at time $t$ & pdf of
waiting times \\
     & from the mother that was born at time $0$ &  \\ \hline
$\Phi(\vec r)$ & pdf for a daughter to be triggered & pdf of jump sizes  \\
& at a distance $\vec r$ from its mother & \\ \hline
$m$ & earthquake magnitude & tag associated with each jump \\ \hline
$\rho(m)$ & number of daughters & local branching ratio \\
& per mother of magnitude $m$  &  \\ \hline
$n$ & average number of daughters created per mother & control
parameter of the random \\
       &  summed over all possible magnitudes &   walk survival
(branching ratio) \\ \hline
$n < 1$ & subcritical aftershock regime & subcritical ``birth and
death'' \\ \hline
$n = 1$ & critical aftershock regime & the standard CTRW  \\ \hline
$n > 1$ & supercritical exponentially & explosive regime of the \\
& growing regime & ``birth and death'' CTRW   \\ \hline
$N(t, \vec r)$ &  number of events of any possible & pdf of just having \\
& magnitude at $\vec r$ at time $t$ & arrived at $\vec r$ at time $t$ \\ \hline
$W(t, \vec r)$ & pdf that an event at $\vec r$ has occurred  at a
time $t' \leq t$
      & pdf of being at $\vec r$ at time $t$ \\
           & and that no event occurred anywhere from $t'$ to $t$  &
\end{tabular}
\vspace{5mm}
\caption{\label{table1}  Correspondence between the ETAS 
(Epidemic-type aftershock sequence)
and CTRW (continuous-time random walk) models. `pdf' stands for 
probability density
function.
}
\end{center}
\end{table}
\newpage

\begin{table}[]
\begin{center}
\begin{tabular}{|l|l|l|}
               & large $z$               & small $z$   \\
               & $r \ll D~t^H$           & $r \gg D~t^H$ \\ \hline
$\mu<0.5$     & $p=1+\theta$            & $p=1-2\theta$\\
$0.5 \leq \mu<2$ & $p=1-\theta+\theta/\mu$ & $p=1-2\theta$ \\
$2 \leq \mu$       & $p=1-\theta/2$          & not defined  \tablenotemark\\
\end{tabular}
\tablenotetext{The Omori exponent is not defined in this case because the
dependence of $N(t,\vec r)$ with respect to time
given by expression (\ref{exp1}) and represented in Figure \ref{nrtmus2} has
a contribution from the exponential asymptotics which is different
from a power-law for large distances $r \gg D~t^H$.}
\vspace{5mm}
\caption{\label{table2}  Asymptotic values of the (renormalized) Omori
exponent (of the dressed propagator)
in the different regimes for $z\ll 1$ and $z\gg 1$ where $z \equiv {
D~t^{H} \over r}$.
}

\end{center}
\end{table}
\newpage

\clearpage

\clearpage

\clearpage

\clearpage

\clearpage

\clearpage

\clearpage

\clearpage
\clearpage

\end{document}